\RequirePackage{fix-cm}
\documentclass[smallcondensed]{svjour3}     % onecolumn (ditto)
\smartqed  % flush right qed marks, e.g. at end of proof
\usepackage{graphicx}
\usepackage{booktabs}
\usepackage{color}
\usepackage{tabularx}
\usepackage{colortbl}
\usepackage[multiple]{footmisc}
\usepackage{adjustbox}
\usepackage{array}
\usepackage{natbib}
\usepackage{enumitem}
\usepackage{blindtext}
\usepackage{scrextend}
\usepackage{amsmath}
\usepackage{tcolorbox}
\usepackage{csquotes}
\usepackage{pdflscape}

\definecolor{Gray}{gray}{0.95}

\newcolumntype{R}[2]{%
    >{\adjustbox{angle=#1,lap=\width-(#2)}\bgroup}%
    l%
    <{\egroup}%
}

\journalname{Software Engineering for Mobile Applications}
\begin{document}

\title{Towards Understanding and Detecting Fake Reviews in App Stores}

\author{Daniel Martens \and Walid Maalej}
%\href{https://orcid.org/0000-0003-0659-5482}{\includegraphics[scale=1]{fig/orcid_16x16}}

\institute{
    Daniel Martens \and Walid Maalej \at
    Department of Informatics\\
    University of Hamburg\\
    Hamburg, Germany\\
    \email{\{martens, maalej\}@informatik.uni-hamburg.de}
}

\date{Received: date / Accepted: date}

\maketitle

\begin{abstract}
App stores include an increasing amount of user feedback in form of app ratings and reviews. 
Research and recently also tool vendors have proposed analytics and data mining solutions to leverage this feedback to developers and analysts, e.g., for supporting release decisions. 
Research also showed that positive feedback improves apps' downloads and sales figures and thus their success. 
As a side effect, a market for fake, incentivized app reviews emerged with yet unclear consequences for developers, app users, and app store operators. 
This paper studies fake reviews, their providers, characteristics, and how well they can be automatically detected. 
We conducted disguised questionnaires with 43 fake review providers and studied their review policies to understand their strategies and offers. 
By comparing 60,000 fake reviews with 62 million reviews from the Apple App Store we found significant differences, e.g., between the corresponding apps, reviewers, rating distribution, and frequency. 
This inspired the development of a simple classifier to automatically detect fake reviews in app stores. 
On a labelled and imbalanced dataset including one-tenth of fake reviews, as reported in other domains, our classifier achieved a recall of 91\% and an AUC/ROC value of 98\%.
We discuss our findings and their impact on software engineering, app users, and app store operators.
\keywords{fake reviews \and app reviews \and user feedback \and app stores}
\end{abstract}

%%%%%%%%%%%%%%%%%%%%%%%%%%%%%%%%%%%%%%%%%%%%%%%%%%%%%%%%%%%%%%%%%%%
\section{Introduction}

In app stores, users can rate downloaded apps on a scale from 1 to 5 stars and write a review  message. 
Thereby, they can express satisfaction or dissatisfaction, report bugs, or suggest new features \citep{6606604, Pagano:2013jn, Maalej:2016:Automatic}. 
Similar to other online stores, before downloading an app, users often read through the reviews.
Research found that ratings and reviews correlate with sales and download ranks \citep{Harman:2012gw, Pagano:2013jn, Svedic:aG1qEZuT, Martin:2016fe, FINKELSTEIN2017119}. Stable numerous ratings lead to higher downloads and sales numbers.

As a side effect, an illegal market for fake app reviews has emerged, with the goal to offer services that help app vendors improve their ratings and ranking in app stores.
According to app store operators, in  regular app reviews, real users are supposed to be triggered by their satisfaction or dissatisfaction of using the app to provide  feedback. Fake reviewers, however, get paid or similarly rewarded to submit reviews. They might or might not be real users of the app. Their review might or might not be correct and reflecting their opinion. 

We refer to this type of non-spontaneous, requested, and rewarded reviews as \textbf{fake reviews}. 
Fake reviews are prohibited in popular app stores such as in Google Play \citep{Weblink:10006} or Apple App Store \citep{Weblink:10005}.
For instance, Apple states: \textit{"If we find that you have attempted to manipulate reviews, inflate your chart rankings with paid, incentivized, filtered, or fake feedback, or engage with third party services to do so on your behalf, we will take steps to preserve the integrity of the App Store, which may include expelling you from the Developer Program."}.

Recently, Google highlighted the negative effects of fake reviews in an official statement and explicitly asks developers to not buy and users to not accept payments to provide fake reviews \citep{Weblink:10021}.
Even governmental competition authorities started taking actions against companies using fake reviews to embellish their apps. For instance, the Canadian telecommunication provider Bell was fined \$1.25 million \citep{Weblink:10007} for faking positive reviews to their apps. Vice versa, the CNN app was affected by thousands of negative fake reviews to decrease its rating and ranking within the Apple App Store \citep{Weblink:20001}.

While the phenomena of fake participation (e.g., in form of commenting, reporting or reviewing) is well-known  in domains such as online journalism \citep{Lee:2010:USS:1835449.1835522, DBLP:journals/corr/FerraraVDMF14, 6921650, DBLP:journals/corr/SubrahmanianADK16} or on business and travel portals \citep{Jindal:2008:OSA:1341531.1341560, Ott:2011:FDO:2002472.2002512, feng2012distributional, Mukherjee2013WhatYF}, it remains  understudied in software engineering -- in spite of recent significant research on app store analysis and feedback analytics \citep{Martin:2016fe}.

Fake reviews threaten the integrity of app stores. If real users don't trust the reviews, they probably will refrain from reading and writing reviews themselves. 
This can result into a problem for app store operators, as app reviews is a central concept of the app store ecosystem. 
Fake reviews can have negative implications for app developers and analysts as well. Numerous software and requirements engineering researchers studied app reviews, e.g., to derive useful development information such as bug reports \citep{Khalid:2013:IUC:2486788.2487044, Maalej:2016:Automatic} or to understand and steer the dialogue between users and developers \citep{6624001, Oh:2013:FDI:2468356.2468681, Johann:RE:2017, 7886888}. 
Further, researchers \citep{6606604, Chen:2014:AMI:2568225.2568263, Maalej:2015:Software} and more recently tool vendors \citep{Weblink:10008} suggested tools that derive actionable information for software teams from reviews such as release priorities and app feature co-existence. 
None of these works considers fake reviews and their implications. Negative fake reviews, e.g., by competitors reporting false issues, can lead to confusion and waste of developers' time. Positive fake reviews might also lead to wrong insights about real users needs and requirements. 

\noindent
In this paper, we study fake app reviews, focusing on three research questions:

\begin{enumerate}[label={\textbf{RQ\arabic*}}, leftmargin = 3.5em,labelsep*=1.0em]
    \item \textbf{How and by whom are app ratings and reviews manipulated?}\\ Through online research and an investigative disguised questionnaire, we identified 43 fake review providers and gathered information about their fake reviewing strategies and offers.
    
    \item \textbf{How do fake reviews differ from regular app reviews?}\\ We crawled $\sim$60,000 fake reviews, empirically analyzed and compared them with $\sim$62 million official app reviews from the Apple App Store. We report on quantitative differences of fake reviewers and concerned apps.
    
    \item \textbf{How accurate can fake reviews be automatically detected?}\\ We developed a supervised classifier to detect fake reviews. Within an in-the-wild experiment, we evaluated the performance of multiple classification algorithms, configurations, and classification features.
\end{enumerate}

In Section~\ref{sec:design} we introduce the research questions, method, and data. 
We then report on the results along the research questions: fake review market in Section~\ref{sec:strategies}, characteristics in Section~\ref{sec:char}, and automated detection in Section~\ref{sec:detection}.
Afterwards, we discuss the implications and limitations of our findings in Section~\ref{sec:discussion}. Finally, we survey related work in Section~\ref{sec:relwork} and conclude the paper in Section~\ref{sec:conclusion}.

% ------------------------------------------------------------------------------
% ------------------------------------------------------------------------------

\section{Study Design}
\label{sec:design}

We first introduce our research questions. Then, we describe our research method and data along the data collection, preparation, and analysis phase.
\vspace{-0.5em} % To increase readability of the following sections

% ------------------------------------------------------------------------------

\subsection{Research Questions}
We aim to qualitatively and quantitatively understand fake app reviews including their market, characteristics, and potential automated detection.
In the following we detail our research questions by listing the sub-questions we aim to answer.

\begin{enumerate}[label={\textbf{RQ\arabic*}}, leftmargin = 3.5em,labelsep*=1.0em]
    \item \textbf{Fake review market} reveals how app sales and downloads are manipulated and to which conditions. 
We investigate the following questions:

    \begin{enumerate}[topsep=0.35em, align=parleft, labelwidth=7.3em, leftmargin=8.3em]
         \item[\textit{1. Providers:}] By whom are fake reviews offered? What strategies do fake review providers follow?
	    \item[\textit{2. Offers:}] What exact services do fake review providers offer and under which conditions?
	    \item[\textit{3. Policies:}] What are providers policies for submitting fake reviews? Do these reveal indicators to detect fake reviews?
    \end{enumerate}

    \item \textbf{Fake review characteristics} reveal empirical differences between official and fake reviews, including reviewed apps and reviewers.

    \begin{enumerate}[topsep=0.35em, align=parleft, labelwidth=7.3em, leftmargin=8.3em]
        \item[\textit{1. Apps:}] Which apps are typically affected by fake reviews? What are their categories, prices, and deletion ratio?
    	\item[\textit{2. Reviewers:}] What is a typical fake reviewer, e.g., in terms of number of reviews provided and review frequency?
    	\item[\textit{3. Reviews:}] How do official and fake reviews differ, e.g., with regard to rating, length, votes, submission date, and content?
    \end{enumerate}

    \item \textbf{Fake review detection} examines how well supervised machine learning algorithms can detect fake reviews. We focus on the following questions:
    
    \begin{enumerate}[topsep=0.35em, align=parleft, labelwidth=7.3em, leftmargin=8.3em]
        \item[\textit{1. Features:}] Which machine learning features can be used to automatically detect fake reviews? 
    	\item[\textit{2. Classification:}] Which machine learning algorithms perform best to classify reviews as fake/non-fake?
    	\item[\textit{3. Optimization:}] How can classifiers further be optimized? What is the relative importance of the classification features?
    	\item[\textit{4. In-the-Wild Experiment:}] How do the classifiers perform in practice on imbalanced datasets with different proportional distributions of fake and regular reviews?
    \end{enumerate}
\end{enumerate}

% ------------------------------------------------------------------------------

\subsection{Research Method and Data}

\begin{figure}[b]
\includegraphics[width=\columnwidth]{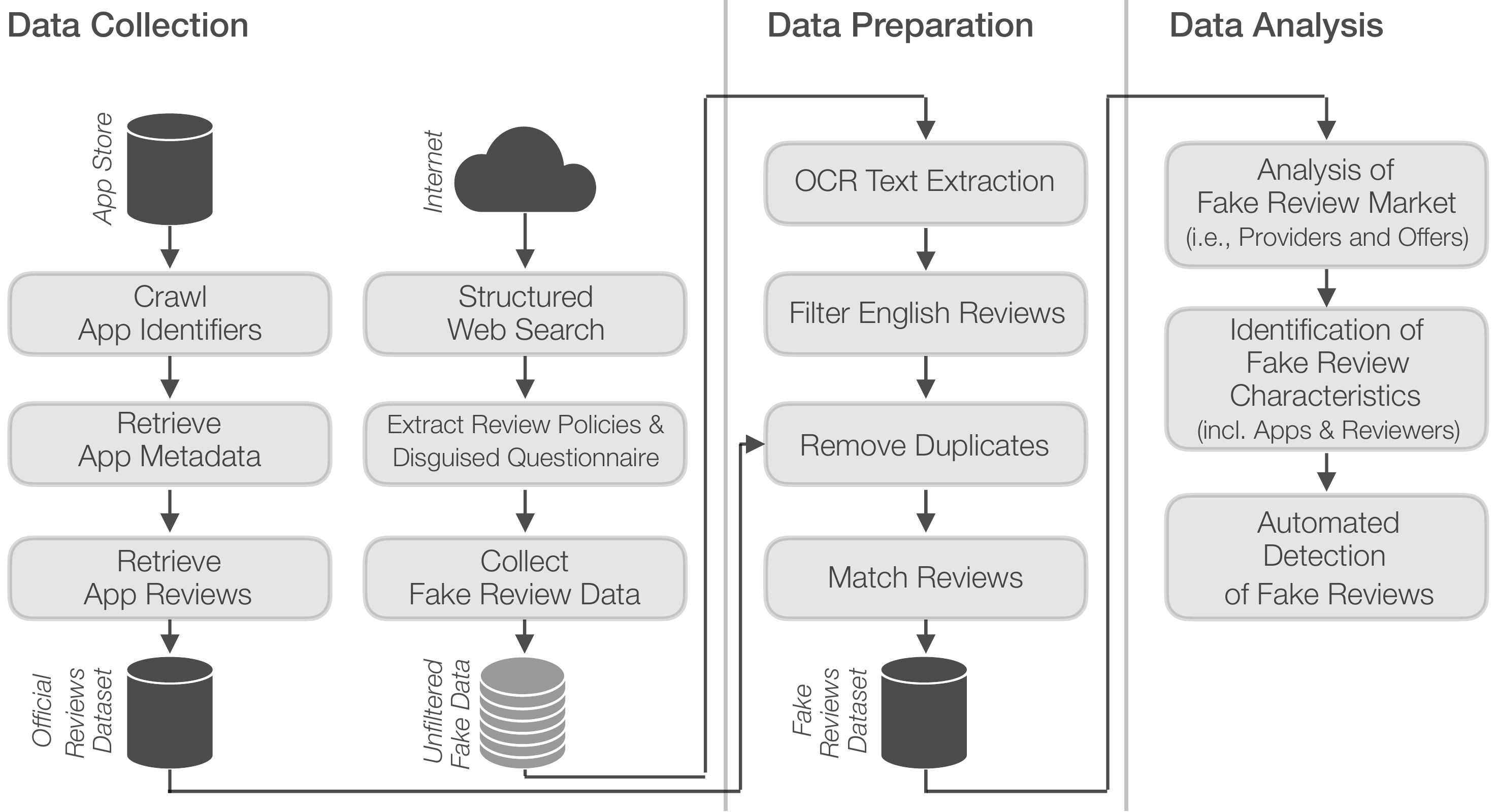}
\caption{Research method including data collection, preparation, and analysis phases.}
\label{fig:researchmethod}
\end{figure}

Our research method consists of a data collection, preparation, and analysis phase, as depicted in Figure \ref{fig:researchmethod}. We detail on each of the three phases in the following.

\subsubsection{Data Collection Phase}

% Overview of Datasets
For this study we collected two datasets: an official reviews dataset including app metadata and reviews from the Apple App Store; as well as a fake reviews dataset including metadata of apps affected by fake reviews, and fake reviews itself.

% App Store Dataset
The \textbf{official reviews dataset} created in March 2017 consists of 1,430,091 apps, their metadata, and reviews. 
To collect the data, we implemented a distributed crawling tool using GoLang based on iTunes APIs, which we deployed on hundreds of cloud servers. 
The data collection included three steps. 
First, we crawled a list of all app identifiers available on the Apple App Store of the United States, as we focus on English reviews.
Second, we obtained the metadata for each app, including the category, price, and number of reviews, using the iTunes Search API\footnote{https://affiliate.itunes.apple.com/resources/documentation/itunes-store-web-service-search-api/}.
Last, we retrieved the app reviews using an internal iTunes API.

Overall, the Apple App Store included 207,782,199 ratings of which 67,727,914 (24.6\%) have a review. 
We were able to crawl 62,617,037 (92.4\%) of these reviews, as iTunes does not allow to receive more than 30,000 reviews per app.
The size of the dataset is 36.58 GB.
The crawled reviews were written by 25,333,786 distinct reviewers, i.e., users with different Apple IDs\footnote{https://appleid.apple.com/faq/\#!\&page=faq}.
On average every reviewer submits around 2.47 reviews.
The oldest app review was entered on 10/07/2008, therefore our dataset spans for nearly 9 years.

\begin{figure}
\centering
\includegraphics[width=.8\columnwidth]{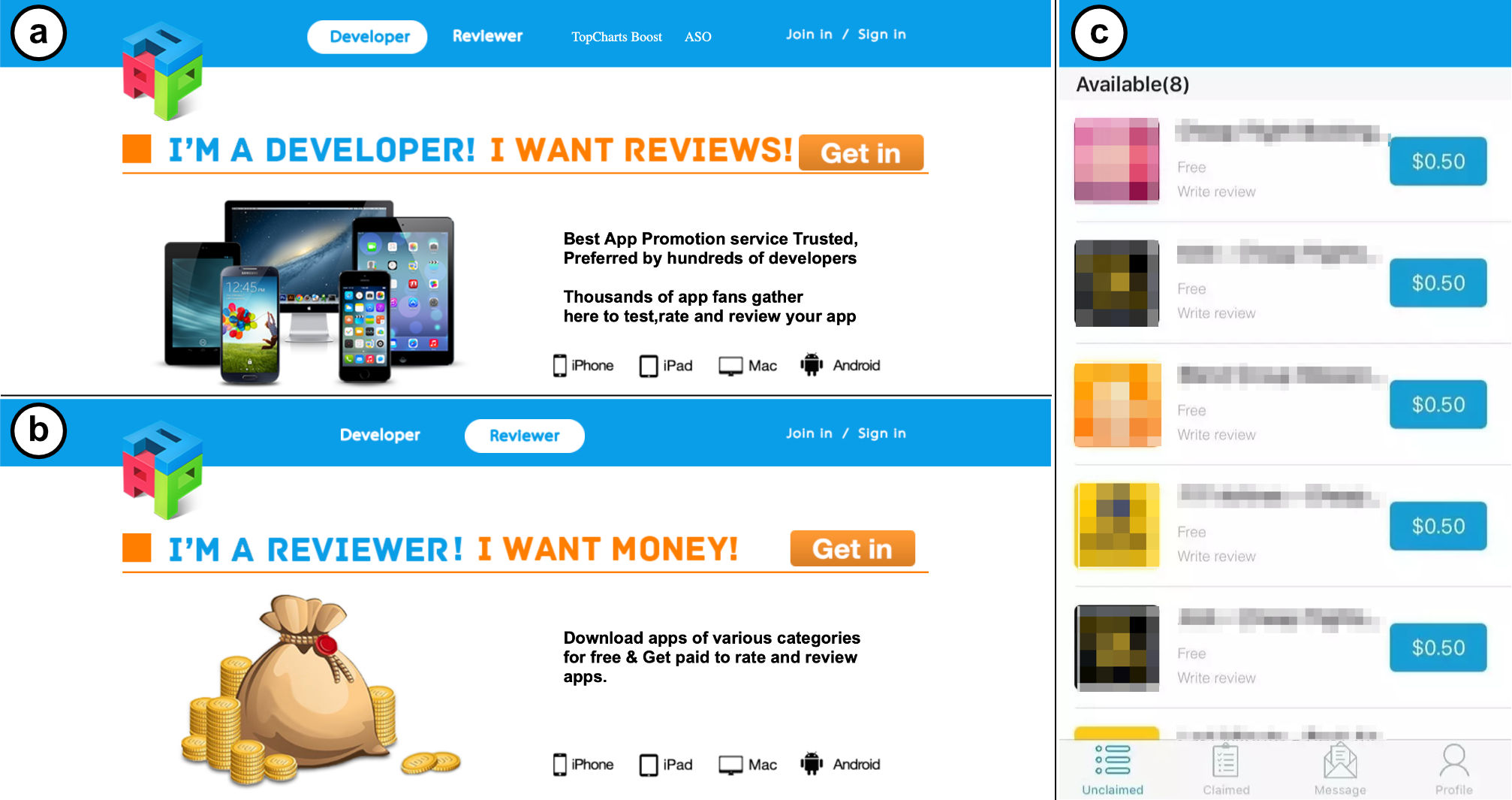}
\caption{Screenshot of a fake review provider, offering a) app developers to buy reviews and b) non-/developers to sign-up and write rewarded reviews. Section c) shows a list of apps requesting fake reviews against monetary reward (obfuscated by the authors).}
\label{fig:providermulti}
\end{figure}

% Fake Dataset
The \textbf{fake reviews dataset} was collected in April 2017 following three steps. 
First, we identified 43 fake review providers by performing a structured manual Google web search.
To identify relevant search terms, we initially searched for the phrase ``buy app reviews''.
We extracted related search terms suggested by the search engine.
For those we repeated the previous step, resulting in 39 unique search terms, which are included in the replication package. 
Afterwards, we crawled the results of the ten first pages for each search term.
We removed duplicate results and marked each result as  fake review provider, relevant discussion about fake reviews (e.g., in forums), or irrelevant result.
From relevant discussions we extracted additional fake review providers by reading through all sub-pages of the discussions.
Then, we manually extracted the provider's offers from their websites. 

In the second step, we conducted a disguised questionnaire to collect initial indicators for fake reviews such as the minimum star-rating and length.  
The questionnaire was presented as a request for buying fake reviews and sent to all providers per email on 26/04/2017\footnote{Conducted with permission of the Ethics Committee of the University of Hamburg.}. 
For providers offering users to sign-up as fake reviewers to exchange or get paid for providing fake reviews, we created accounts and extracted the policies submitted fake reviews must comply with (cf. Figure~\ref{fig:providermulti}).

\begin{table}%[b]
\centering
\renewcommand{\arraystretch}{1.2}
\caption{Overview of collected fake reviews and apps affected per provider (extracted using the listed approaches, dashes indicate that no apps or reviews could be extracted).}
\label{tab:fakereviewdataset}
\begin{tabularx}{\columnwidth}{XlrrX}
\hline\noalign{\smallskip}
Provider Id   & Provider Type & \# Apps        & \# Reviews     & Approach     \\
\noalign{\smallskip}\hline\noalign{\smallskip}
\rowcolor{Gray} PRP10 & Paid Review Provider      & 77            & -              & Crawl         \\
PRP16 & Paid Review Provider      & 19            & 4              & Crawl, Social \\
\rowcolor{Gray} PRP21 & Paid Review Provider      & 3             & -              & Social        \\
PRP25 & Paid Review Provider      & -             & 3              & Social        \\
\rowcolor{Gray} PRP26 & Paid Review Provider      & -             & 10             & Social        \\
PRP28 & Paid Review Provider      & -             & 3              & Social        \\
\noalign{\smallskip}\hline\noalign{\smallskip}
\rowcolor{Gray} REP1 & Review Exchange Portal & 268           & -              & Crawl         \\
REP2 & Review Exchange Portal & 277           & -              & Crawl         \\
\rowcolor{Gray} REP3 & Review Exchange Portal & 2,007         & 60,411         & API, Crawl    \\
REP5 & Review Exchange Portal & 7             & -              & Crawl         \\
\rowcolor{Gray} REP6 & Review Exchange Portal & 9             & -              & Crawl         \\
REP8 & Review Exchange Portal & 182           & -              & Crawl         \\
\rowcolor{Gray} REP9 & Review Exchange Portal & 4             & -              & Crawl         \\
\noalign{\smallskip}\hline\noalign{\smallskip}
    &          & $\sum$ = 2,853 & $\sum$ = 60,431 &               \\ 
\noalign{\smallskip}\hline
\end{tabularx}
\end{table}

In the third and last step, we collected the fake review data, i.e., lists of apps requesting fake reviews and fake reviews itself. 
For this, we used three approaches: 

\begin{enumerate}
\item \textit{Social investigation}, i.e., asking the providers for fake review examples, while pretending to be interested in their services. 
We contacted the providers via email or live-chats on their websites.
Using this strategy we received 3 apps and 20 fake reviews from 5 providers.
\item \textit{Crawling}, for providers offering to sign-up as fake reviewers, we checked if the lists of apps requesting fake reviews are available (see  Figure~\ref{fig:providermulti}, part c).
To extract the apps we implemented crawlers. A sample crawler is included in the replication package.
Overall, we collected 2,850 apps from 9 providers.
\item \textit{APIs}, we found that providers require reviewers to upload screenshots of their reviews as a proof. We searched for publicly accessible screenshots and downloaded them. 
Based on this, we gathered 60,411 reviews from a single provider.
\end{enumerate}

\textbf{Overall we identified 2,853 apps and 60,431 reviews} from 13 providers, see Table~\ref{tab:fakereviewdataset}.
Per provider the number of extracted apps and reviews is given, in case we could successfully apply at least one of the introduced approaches.
The size of the collected data is 11.29 GB. 
We refer to this as \textbf{unfiltered fake data} (cf. Figure~\ref{fig:researchmethod}), as it needs to be prepared for further analysis. 
For example, reviews within the dataset could have already been removed from the app store. 
For data preparation and analysis, all data, except the screenshots, is persisted as Parquet files and analyzed with Apache Zeppelin and Spark\footnote{https://zeppelin.apache.org/}.

\subsubsection{Data Preparation Phase}

\begin{figure}
\centering
\includegraphics[width=.8\columnwidth]{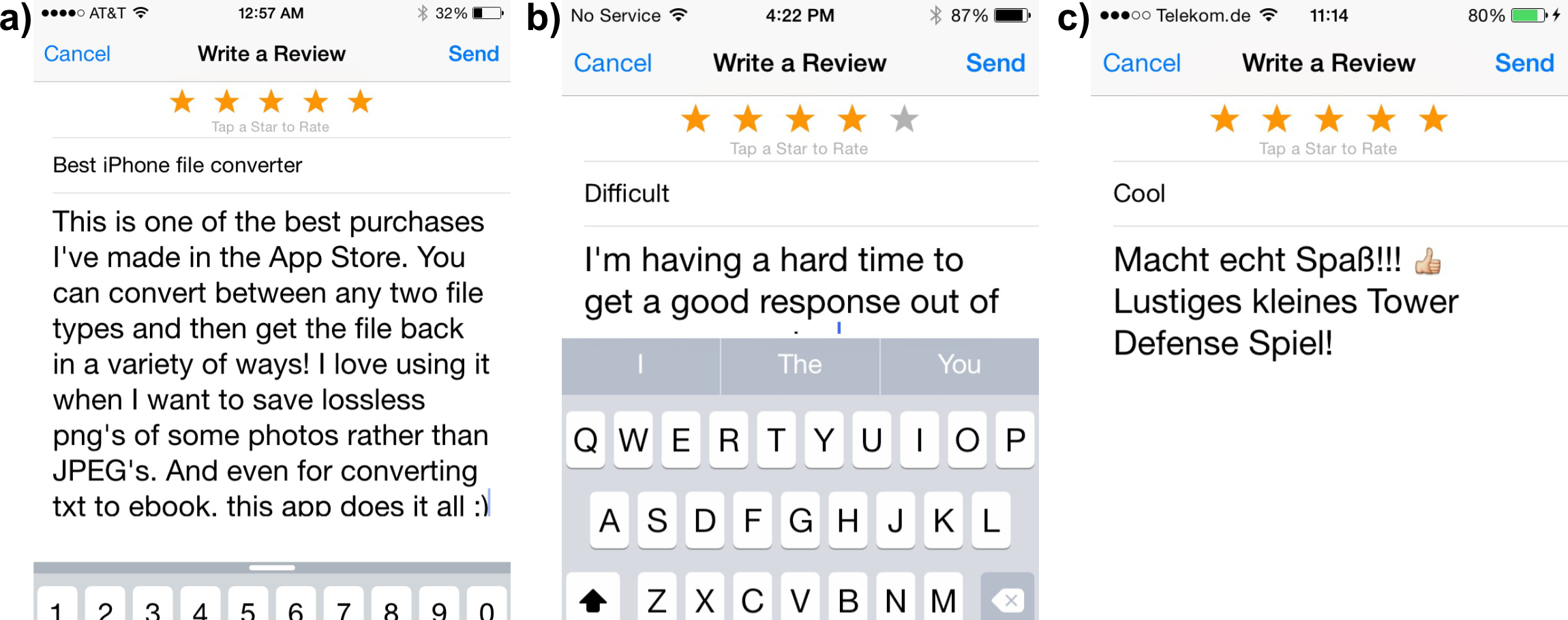}
\caption{Screenshots of fake reviews before submission to the app store used as proof for fake review providers, depicting a) fake review included in our study, b) cut-off fake review excluded from the study, and c) non-English fake review also  excluded from the study.}
\label{fig:reviewscreen}
\end{figure}

Most fake reviews were collected in form of screenshots as shown in  Figure~\ref{fig:reviewscreen}.
We converted the screenshots into text using the Tesseract OCR engine\footnote{https://github.com/tesseract-ocr/tesseract}. 
We removed incomplete reviews that do not include a full readable title and body, e.g., if the title was outside the screen's visible area.
Then, we used the Language Identification (LangID) library\footnote{https://github.com/saffsd/langid.py} to retrieve fake reviews in English language only. 
We removed 7,445 reviews (12.32\%) resulting in 52,986 fake reviews in English language.

Since the screenshots show the review edit screens before submitting the reviews to the app store, we further filtered the collected fake reviews for three possible reasons. 
First, we cannot assure that the reviews were actually submitted. 
Second, the reviews could have not been unlocked by the app store operator.
Third, the reviews could have been deleted.
Therefore, we only considered fake reviews that have been published to and still exist within the Apple App Store, i.e., which we could identify in the official reviews dataset as well.

For uniquely identifying (i.e., matching) reviews from the fake reviews dataset within the official reviews dataset, we removed duplicate reviews which consist of the same title and body within both datasets.
Thereby, we removed 4,298 (8.11\%) fake reviews leaving 48,688 items. 
The percentage of duplicate reviews within the official reviews dataset is with 16.08\% nearly twice as high, which may be an indicator for the high diversity of fake reviews.
We performed the matching using exact text comparison and by comparing the reviews' Levensthein distances. 
We used the Levensthein distance as single characters on the screenshots were sometimes not parsed correctly by the OCR engine.
We searched for all fake reviews within the official reviews dataset by using an edit distance of up to 10 characters.
For possible matches identified using the Levensthein distance, we manually verified if one of the suggested pairs is a match using two human annotators comparing the screenshot of the fake review and the possible matches. In case of disagreements (3\% of all cases), a third annotator resolved the conflict. 
We matched 6,020 reviews by exact text comparison and 2,584 reviews by comparing the Levensthein distance.

Overall, we were able to identify 8,607 of the 60,431 (14.2\%) collected fake reviews within the Apple App Store. 
These reviews were extracted from 5 providers.
%
% Apps
We also matched \textit{apps} affected by fake reviews against the official reviews dataset, as the apps might not be available in the US App Store or might have been deleted.
Of the 2,853 collected apps we found 2,174 apps (76.2\%) in the official reviews dataset. 
Further, we identified 898 additional apps by extracting the app identifiers from previously matched fake reviews, resulting in 3,072 apps.
We removed all apps that did not receive reviews within the app store, resulting in 1,929 of 3,072 (62.8\%) apps provided by 10 different providers.
%
% Users
Finally, we identified 721 fake \textit{reviewers}, i.e., accounts of persons submitting fake reviews to the app store, by extracting their user identifiers from fake reviews.

In summary, after data cleaning the \textbf{fake reviews dataset} consists of 43 providers and structural information about their offers and policies, as well as \textbf{8,607 fake reviews, 1,929 apps affected by fake reviews, and 721 fake reviewers}.
%The size of the dataset after text extraction from the screenshots and data cleaning is 34 MB. 
The dataset spans for nearly 7 years, as the oldest fake review was entered on 16/10/2010.
Table~\ref{tab:datasets} summarizes the official and fake reviews datasets.

\begin{table}
\centering
\renewcommand{\arraystretch}{1.3}
\caption{Overview of the official and the fake reviews datasets.}
\label{tab:datasets}
\begin{tabularx}{\columnwidth}{Xrr}
\hline\noalign{\smallskip}
 & Official Reviews Dataset & Fake Reviews Dataset \\
\noalign{\smallskip}\hline\noalign{\smallskip}
\rowcolor{Gray}\# of reviews & 62,617,037 & 8,607 \\
\# of apps & 1,430,091 & 1,929 \\
\rowcolor{Gray}\# of reviewers & 25,333,786 & 721 \\
\noalign{\smallskip}\hline
\end{tabularx}
\end{table}

\subsubsection{Data Analysis Phase}

The data analysis phase consists of three steps which respectively answer the research questions. 
% Market
To answer our first research question regarding the fake review \textbf{market}, we analyzed the qualitative data aggregated from the providers' websites, collected during the questionnaire, and extracted from the review policies. 

% Characteristics
To explore the fake review \textbf{characteristics} we applied a statistical analysis of the reviews, apps, and reviewers. We compare the figures from the fake reviews dataset to those from the official reviews dataset and run statistical tests whenever applicable. 
For example, we found that most fake reviews are provided for games. 
While regular apps receive most reviews on the day of an app release, apps affected by fake reviews receive most reviews eleven days after the update. 
This could indicate that apps affected by fake reviews do not have a real users basis that intrinsically provides reviews in reaction to changes introduced by app updates.
Further, we found that fake reviewers provide twelve times as much reviews, with a four times higher frequency. 
Also, fake reviews have more positive ratings compared to official reviews, however the biggest difference exists between the amount of 1-star ratings. 

% Classifier
To \textbf{detect} fake reviews, we created a labeled and balanced dataset of fake reviews and official reviews and used it to train and evaluate multiple classifiers, based on machine learning features that we derived from the analysis of characteristics. 
We conducted a hyperparameter tuning of the classifiers and evaluated the importance of the classification features. 
Finally, we performed an in-the-wild experiment to get more realistic results of how the classifiers perform in practice.
Therefore, we used imbalanced datasets of fake and regular reviews. We varied the skewness of the datasets between 90\% to 0.1\% fake reviews and compared the classification results.

% Summary
We detail each of these analysis steps in the following chapters. 
% Replication
To support replication, our dataset and the analyses source code as Zeppelin notebooks are publicly available on our website\footnote{https://mast.informatik.uni-hamburg.de/app-review-analysis/}.

% ------------------------------------------------------------------------------
% ------------------------------------------------------------------------------

\section{Fake Review Market (RQ1)}
\label{sec:strategies}

This section describes fake review providers and their market strategies, as well as offers and pricing models.  
Afterwards, pretended characteristics of fake reviews are summarized based on the results of the disguised questionnaire and analysis of reviewing policies.

% ------------------------------------------------------------------------------

\subsection{Review Providers and Market Strategies}

We identified 43 providers offering fake reviews. These can be separated into two groups by their strategies used to supply reviews. 

\begin{figure}
\centering
\includegraphics[width=\columnwidth]{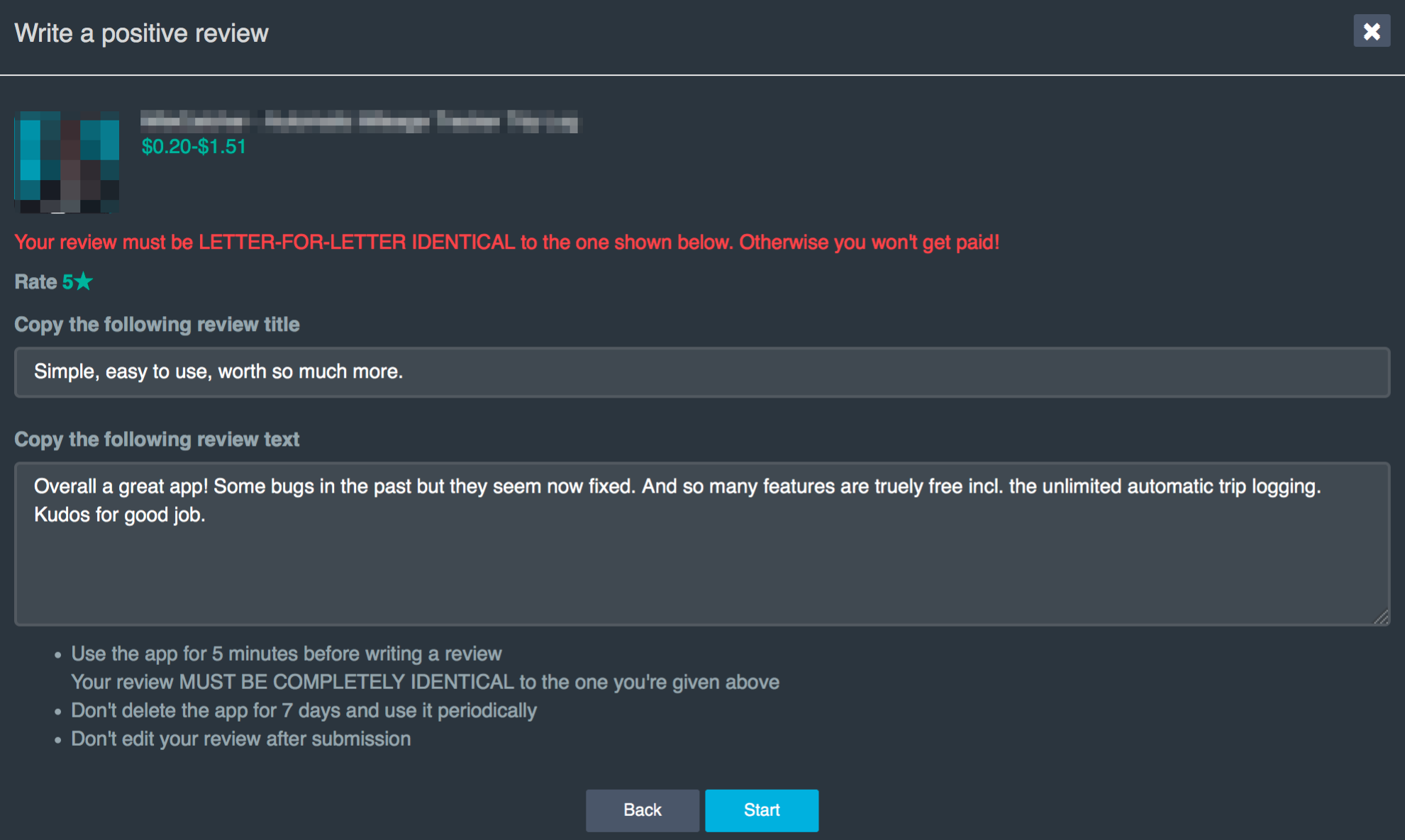}
\caption{A fake review request with predefined rating and review on a review exchange portal.}
\label{fig:reviewrequest}
\end{figure}

\textbf{Paid review providers (PRP) accept payments to provide fake reviews.} 
This applies for 34 out of 43 (79\%) providers. 
User can select a package of, e.g., 50 reviews, specify their app name and identifier, and purchase it via Paypal or similar services.
Afterwards, the fake reviews are submitted to the app store.

\textbf{Review exchange portals (REP) allow app developers to sign-up and exchange reviews.}
The remaining 9 providers (21\%) belong to this group. 
After sign-up developers browse through a list of apps requesting fake reviews.
Figure~\ref{fig:reviewrequest} shows a sample  request for fake reviews.
Depending on their policies, review exchange portals ask users to submit fake reviews, e.g., with predefined ratings and review messages.
For each fake review the developer submits, one credit is given as a reward. 
Developers with at least one credit can add their app to the list.
Then, the credits are redeemed into reviews written by other developers.

In some cases, review exchange portals allow developers to buy credits and non-developers to sign-up and submit fake reviews. 
Non-developers are rewarded using micro-payments, typically between \$0.20 to \$1.50 per fake review.

\begin{figure}[t]
\centering
\includegraphics[width=\columnwidth]{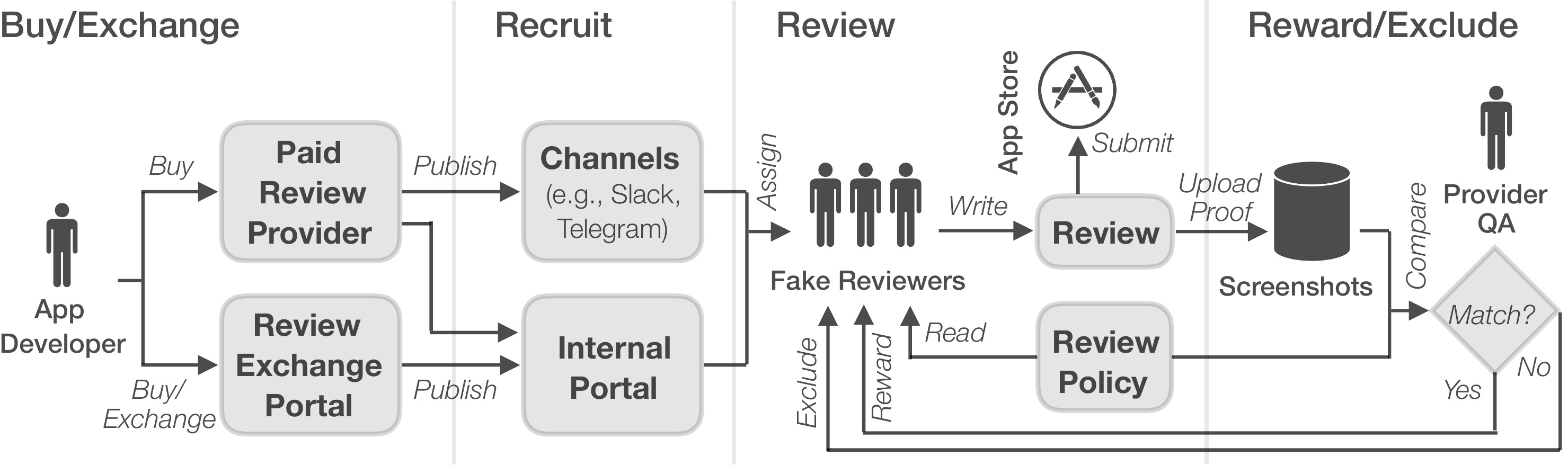}
\caption{Fake reviewing strategies.}
\label{fig:reviewingstrategy}
\end{figure}

Figure~\ref{fig:reviewingstrategy} shows the strategies of the fake review providers.
After deciding to buy fake reviews at a paid review provider or to exchange (or buy) reviews at a review exchange portal, the developer provides basic information, such as the application identifier and whether the reviews should be positive or negative. Optionally, further information, such as keywords to be included within the reviews or predefined review messages, can be submitted.
Using this information, the provider creates a review request (see, e.g., Figure~\ref{fig:reviewrequest}).

Review exchange portals publish these requests on their internal platform to recruit fake reviewers.
For paid review providers the publishing process is not transparent.
Using social investigation and by offering our service as fake reviewer, we identified that at least five providers publish their review requests on invite-only Slack or Telegram channels.
By observing the communication within these channels, we found that paid review providers occasionally cross-post review requests on review exchange portals while offering micro payments.

Fake reviewers can browse through and assign review requests to themselves.
Afterwards, they are presented a review policy that regulates what information or rating the review should include.
The fake reviewer submits an appropriate fake review to the app store. 
As a proof, the fake reviewer uploads a screenshot of the review edit screen showing their rating and review to the provider.

Last, the provider compares if the provided review meets the reviewing policy.
If this applies, and the review has been published within the app store, the fake reviewer is rewarded.
Reviewers providing reviews that do not meet the policies are excluded from the channels or portals and are not rewarded.

\begin{table}
\centering
\renewcommand{\arraystretch}{1.3}
\caption{Offers and prices (in US\$) of paid review providers.}
\label{tab:paidprovider}
\begin{tabularx}{\columnwidth}{llXXXXXXXXXXXX}
\hline\noalign{\smallskip}
PRP & Co. & \multicolumn{4}{l}{Review Price}                           & \multicolumn{4}{l}{Rating Price}                           & \multicolumn{4}{l}{Install Price}                          \\ \noalign{\smallskip}\cline{3-14}\noalign{\smallskip} 
         &         & \multicolumn{2}{l}{iOS} & \multicolumn{2}{l}{Android} & \multicolumn{2}{l}{iOS} & \multicolumn{2}{l}{Android} & \multicolumn{2}{l}{iOS} & \multicolumn{2}{l}{Android} \\ \noalign{\smallskip}\cline{3-14}\noalign{\smallskip} 
         &         & Min  & Max  & Min    & Max    & Min  & Max  & Min    & Max    & Min  & Max  & Min    & Max    \\ 
\noalign{\smallskip}\hline\noalign{\smallskip}
\rowcolor{Gray}1 &IN & & & 1.35 & 1.50 & & & & & & & &  \\ 
2 &DK & 4.63 & 4.90 & 0.98 & 1.00 & & & & & 0.25 & 0.25 & 0.05 & 0.06\\
\rowcolor{Gray}3 &IN & & & 0.25 & 0.25 & & & 0.20 & 0.20 & & & 0.09 & 0.09\\
4 &IN & & & 1.50 & 1.98 & & & & & & & &  \\
\rowcolor{Gray}5 &GB & & & 1.11 & 1.50 & & & & & & & 0.10 & 0.10\\
6 &US & & & 2.90 & 2.95 & & & 1.28 & 1.58 & & & 1.30 & 1.36\\
\rowcolor{Gray}7 &RU & & & 0.25 & 0.25 & & & 0.20 & 0.20 & & & 0.10 & 0.10\\
8 &US & 6.00 & 9.00 & \textbf{6.00} & \textbf{9.00} & & & & & & & &  \\
\rowcolor{Gray}9 &US & 3.33 & 4.17 & 1.00 & 1.50 & & & & & & & 0.09 & 0.15\\
10 &NL & & & 1.55 & 1.55 & & & 1.00 & 1.00 & 0.65 & 0.65 & 0.20 & 0.20\\
\rowcolor{Gray}11 &US & 2.50 & 4.00 & 3.50 & 5.00 & & & & & 0.49 & 0.90 & 0.08 & 0.12\\
12 &CA & & & 1.00 & 1.00 & & & & & & & &  \\
\rowcolor{Gray}13 &US & 2.15 & 2.50 & 1.59 & 2.50 & & & & & 0.34 & 0.46 & 0.13 & 0.20\\
14 &US & 4.30 & 5.00 & 0.85 & 1.20 & & & & & 0.35 & 0.38 & 0.35 & 0.38\\
\rowcolor{Gray}15 &IN & & & 0.15 & 0.15 & & & 0.08 & 0.08 & & & 0.10 & 0.10\\
16 &RU & 2.09 & 2.99 & 2.99 & 2.99 & & & & & & & &  \\
\rowcolor{Gray}17 &US & 5.02 & 5.20 & 2.00 & 2.60 & & & & & 0.40 & 0.45 & 0.40 & 0.46\\
18 &DE & & & 2.50 & 2.50 & & & & & & & 0.17 & 0.17\\
\rowcolor{Gray}19 &US & \textbf{8.69} & \textbf{10.00} & 3.60 & 4.00 & & & 1.28 & 1.60 & & & 1.36 & 1.58\\
20 &VN & & & 0.05 & 0.05 & & & 0.05 & 0.05 & 0.10 & 0.10 & 0.05 & 0.05\\
\rowcolor{Gray}21 &US & 2.00 & 2.00 & 1.40 & 2.00 & & & & & & & &  \\ 
22 &US & & & 1.45 & 2.00 & & & & & 0.29 & 0.32 & 0.08 & 0.15\\
\rowcolor{Gray}23 &RU & 3.40 & 4.00 & 2.75 & 2.75 & & & & & & & &  \\ 
24 &US & & & 1.00 & 1.00 & & & 0.80 & 0.80 & & & 0.15 & 0.15\\
\rowcolor{Gray}25 &NL & 1.78 & 3.30 & 1.78 & 3.30 & & & & & 0.50 & 0.50 & 0.08 & 0.10\\
26 &RU & 3.00 & 3.00 & & & & & & & & & &  \\
\rowcolor{Gray}27 &IN & & & 1.99 & 2.40 & & & & & 0.39 & 0.46 & 0.39 & 0.46\\
28 &CN & 2.09 & 2.99 & 2.39 & 2.99 & 1.00 & 1.99 & & & & & &  \\
\rowcolor{Gray}29 &SG & 3.00 & 3.00 & 3.00 & 3.00 & & & & & & & &  \\ 
30 &DE & 1.93 & 4.00 & & & & & & & 0.45 & 0.50 & 0.06 & 0.14\\
\rowcolor{Gray}31 &US & & & 1.00 & 2.00 & & & & & & & 0.50 & 1.00\\
32 &IN & & & 0.50 & 0.50 & & & & & & & &  \\
\rowcolor{Gray}33 &AE & & & 0.90 & 1.00 & & & 0.75 & 0.80 & & & 0.15 & 0.40\\
34 &IN & 2.00 & 2.00 & 2.00 & 2.00 & \textbf{2.00} & \textbf{2.00} & \textbf{2.00} & \textbf{2.00} & \textbf{1.60} & \textbf{1.67} & \textbf{1.60} & \textbf{1.67}\\
\noalign{\smallskip}\hline\noalign{\smallskip}
\rowcolor{Gray} NUM     &     & 17 & 17      & 32 & 32     & 2 & 2       & 10 & 10      & 12 & 12     & 23 & 23\\
AVG   &     & 3.41 & 4.24 & 1.73 & 2.14 & 1.50 & 2.00 & 0.76 & 0.83 & 0.48 & 0.55 & 0.33 & 0.40\\
\rowcolor{Gray} SD &     & 1.85 & 2.21 & 1.24 & 1.70 & 0.71 & 0.01 & 0.64 & 0.71 & 0.38 & 0.40 & 0.45 & 0.50\\
\noalign{\smallskip}\hline
\end{tabularx}
\end{table}

\subsection{Offers and Pricing Models}

% Offer overview and description of table
To increase app downloads and sales, paid review providers offer fake reviews, ratings, and installs.
Table \ref{tab:paidprovider} shows the prices of these offers for both the Android and iOS platform.
The table lists the offers' minimum and maximum price, which varies, e.g., depending on the amount of reviews bought.

% Reviews
\textbf{Paid fake reviews} are offered by 17 of 34 (50\%) providers for iOS and by 32 of 34 (94.1\%) for Android. 
Reviews always include a rating.
Among all offers, reviews are the most expensive. 
The price of a review for iOS is, on average, between \$3.41 and \$4.24 with a standard deviation from 1.85 to 2.21.
The price of a review for Android is less expensive, on average, between \$1.73 and \$2.14 with a standard deviation from 1.73 to 2.14.
The price for iOS is $\sim$97 to 98\% higher.

% Ratings
\textbf{Paid fake ratings} are offered by two of 34 (5.9\%) providers for iOS and by 10 of 34 (29.4\%) for Android. 
With this offer fake reviewers rate the app without submitting a written review.
The price of a rating for iOS is on average between \$1.50 and \$2.00 with a standard deviation from 0.01 to 0.71.
The price of a rating for Android is on average between \$0.76 to \$0.83 with a standard deviation of 0.76 to 0.83.
Compared to Android the price for iOS ratings is $\sim$97 to 140\% higher.

% Installs
\textbf{Paid installs} are generated, e.g., by advertising the app on blogs.
Also, users can be paid to install the app. 
The acquired new app users decide by themselves to rate and review the app. 
According to our definition these reviews are not considered as fake, as these are not directly requested or paid for. 
Installs are offered by 12 of 34 (35.4\%) providers for iOS and by 23 of 34 (67.6\%) for Android.
Among all offers installs are the least expensive. 
The price of an iOS app install is between \$0.48 to \$0.55 with a standard deviation of 0.38 to 0.40.
For Android the price is between \$0.33 to \$0.40 with a standard deviation of 0.45 to 0.50.
Comparing both platforms the price difference is  $\sim$37 to 45\%.

Overall, the offers are rather expensive, e.g., compared to an average crowdsourcing task or buying followers on Twitter. 10,000 Twitter followers can be bought for a price of \$4 \citep{Stringhini:2013:FGG:2504730.2504731}. This might depend on the fact that fake ratings and reviews are generated manually, e.g., due to a strict moderation by app store operators, while Twitter followers can be generated automatically.

We further tried to identify the popularity of the three different types of offers. 
Unfortunately, we were only able to extract usage numbers from paid review provider 10 (PRP10) and thus cannot provide generalizable information. Overall, this provider sold 354,000 offers, of which 20,750 (5.9\%) were fake reviews, 29,150 (8.2\%) fake ratings, and 304,100 (85.9\%) paid installs.
We cannot give any numbers on how many paid installs result into a rating or review, and if these are comparable to fake ratings and reviews.

% ------------------------------------------------------------------------------

\subsection{Pretended Fake Review Characteristics}

To understand the rules and conditions of providing fake reviews, we conducted a disguised questionnaire with the paid review providers. We also extracted the policies, with which submitted reviews must comply in review exchange portals.

\subsubsection{Disguised Questionnaire}

% Survey with Paid Review Providers
The disguised questionnaire consists of eleven questions and was presented to providers in a request for buying fake reviews. 
A sample question is shown below:

\begin{displayquote}
\textit{We have several competitors which gain more and more market share.
For this reason we are looking for both positive and negative reviews, positive for our apps and negative for our competitors' apps. [...]}
\end{displayquote}

We decided against open questions as we noticed during a pre-run of the questionnaire, conducted using different identities, that providers returned incomplete answers. The questionnaire is included in our replication package.

Eleven out of 34 paid review providers (32.3\%) answered our questionnaire. 
Table \ref{tab:survey} summarizes their answers. 
Even upon request, not all providers answered all of our questions.
Therefore, the total answers refer to the number of providers that explicitly answered the specific question. 

\begin{table}
\centering
\renewcommand{\arraystretch}{1.3}
\caption{Summary of disguised questionnaire showing offers of paid review providers.}
\label{tab:survey}
\begin{tabularx}{\columnwidth}{lXXXXXX}
\hline\noalign{\smallskip}
PRP & Positive Ratings & Negative Ratings & Custom Keywords & Predefined Reviews & Real Users & Guarantee \\
\noalign{\smallskip}\hline\noalign{\smallskip}
\rowcolor{Gray}9                & Yes                & No                 & Yes               & Yes                 & Yes                 & Yes                \\ 
10               & Yes                & Yes                &                  &                    & Yes                 & No                 \\
\rowcolor{Gray}12               & Yes                & Yes                & Yes               & No                  & Yes                 & Yes                 \\ 
15               & Yes                & Yes                &                  &                    &                    & No                 \\
\rowcolor{Gray}16               & Yes                & No                 & Yes               & Yes                 & Yes                 & No                  \\ 
22               & Yes                & Yes                &                  &                    & Yes                 &                   \\
\rowcolor{Gray}23               & Yes                & No                 & No                & Yes                 & Yes                 & No            \\
25               & Yes                & Yes                & Yes               & Yes                 & Yes                 & Yes                 \\
\rowcolor{Gray}26               & Yes                & Yes                & Yes               & Yes                 & Yes                 & Yes            \\
28               & Yes                & No                 & No                & Yes                 & Yes                 & No                 \\
\rowcolor{Gray}29               & Yes                &                   & Yes               & Yes                 & Yes                 & Yes                \\
\noalign{\smallskip}\hline
\end{tabularx}
\end{table}

While all 11 providers offer positive ratings and reviews, 6 also offer negative, e.g., to lower the reputation of competing apps. 
Regarding the content, 6 of 8 providers reported to accept keywords, which will be included in their reviews.
Seven of 8 providers accept predefined reviews to be submitted by their reviewers.
All providers state their reviews are written by humans and not generated using algorithms.
Five of 10 providers gave a guarantee to replace deleted fake reviews. 

Regarding the geographical origin of fake reviews, PRP10 and PRP15 provide reviews from the US. PRP23 and PRP26 additionally provide reviews from Russia. 
P25 also provides reviews from India. PRP28 specified 13 countries from which the reviews are submitted, these are Austria, Canada, China, France, Germany, India, Italy, Japan, Russia, Taiwan, United Kingdom, United States, and Vietnam. 
Four providers (PRP9, PRP12, PRP16, and PRP29) reported to submit reviews from all over the world. 
According to the results, the top three countries reviews are provided from are: United States (54.5\%), Russia (36.4\%), and India (18.2\%).
Regarding the language, PRP9 and PRP29 reported to provide reviews in all languages. 
PRP10, PRP12, PRP15, PRP25, and PRP28 only provide reviews in English. 
PRP23 and PRP26 also provide reviews in Russian language.
By analyzing all 60,431 fake reviews, initially collected, using LangID, we found that these are written in 70 languages. 
The five most common are English (87.7\%), French (2.3\%), German (2.2\%), Italian (1.3\%), and Spanish (1\%).

\subsubsection{Review Policies}

% Review Policies of Review Exchange Portals
For all review exchange portals, we were able to extract policies that state the requirements submitted fake reviews have to confirm to, see Table \ref{tab:guidelines}.
The policies have different levels of details. Thus, not every requirement is stated by each policy.
With the total number we refer to the policies that explicitly state a requirement.

\begin{table}
\centering
\renewcommand{\arraystretch}{1.3}
\caption{Review characteristics extracted from policies of review exchange portals.}
\label{tab:guidelines}
\begin{tabularx}{\columnwidth}{llXXllXXlX}
\hline\noalign{\smallskip}
REP & Co.                        & Real Dev. & Install App & Use App    & Keep App     & Honest & Rating       & Length     & Copy \\
\noalign{\smallskip}\hline\noalign{\smallskip}
\rowcolor{Gray}1 & IN &       & Yes     & No     &    &       & 1-5  &           &     \\ 
2 & ES & Yes    & Yes     &       & 5 days   & Yes    & 3-5  & $>$10 words  & No   \\
\rowcolor{Gray}3 & US & Yes    & Yes     &       & 1-2 days & Yes    &       & 2-3 sentences &     \\ 
4 & US &       &        &       &         &       &       & 1-2 sentences & No   \\
\rowcolor{Gray}5 & GB & Yes    & Yes     &       & 1 day    & Yes    & 4-5  & 1-2 sentences &     \\ 
6 & CN & Yes    & Yes     & 4 min  & 5 days   &       &       &           &     \\
\rowcolor{Gray}7 & GB &       &        &       &         & Yes    &       & 1-2 sentences &     \\ 
8 & SE & Yes    & Yes     &       & 2 days   & Yes    & 3-5  & $>$10 words  & No   \\
\rowcolor{Gray}9 & RU &       & Yes     & 10 min & 7 days   &       &       &           &     \\
\noalign{\smallskip}\hline
\end{tabularx}
\end{table}

Five portals require to use a real device to submit a review. 
The installation of the app is explicitly required by seven portals.
Only two portals request the reviewers to use the app before submitting a review. 
REP6 requires the reviewer to use the app for at least 4 minutes and REP9 for at least 10 minutes. 
REP1 explicitly states that the usage of the app is not required. 
Six portals state that the app should be kept on the phone for a specific amount of time after leaving the review.
The minimum amount of time is one (REP3, REP5) up to 7 days (REP9).

Regarding the rating, four providers specify a range the rating should follow. REP2 and REP8 request 3-5 stars, and REP5 4-5 stars. REP1 is the only provider explicitly allowing positive and negative (1-5 stars) ratings.
However, reviews from REP1 are not included in our fake reviews dataset.
Five providers state that the review should be honest, although three of those allow only positive ratings with at least 3-4 stars. These providers state that reviewers should skip apps if they are unable to submit a positive review.

Regarding the review length, six portals make a statement: two require at least 10 words, three portals 1-2 sentences, and one portal 2-3 sentences.
Three providers explicitly state that the review should not contain content copied from the app description.
REP5 and REP9 additionally require that the reviews are sufficiently detailed, e.g., ``should describe app features instead of providing only praise''.

REP2 allows reviewers to only rate up to 10 apps per day.
Further, the app should not be immediately reviewed after its installation.
Reviewers should randomly download apps from the app store without leaving a review.
Before leaving another review the reviewer should wait a few minutes.
Last, reviewers should not only provide 5-star ratings, but vary between 3-5 star ratings.
REP7 requires the ratings to match review content.
Finally, REP9 requires to launch reviewed apps periodically within the next 7 days. A possible reason for this, might be to hide the suspicious behavior of quitting to use an app after providing a positive review from app store operators.

\subsubsection{Initial Fake Review Indicators}

Considering the questionnaire results, the review policies, and taking into account the efforts of providers to disguise fake reviews, we can hypothesize that fake reviews are highly diverse. For example, the \textit{rating} of fake reviews can be positive or negative. The \textit{length} of a review can also vary. In addition, the quality of the \textit{content} may strongly differ. Overall, fake reviews do not mean short and low quality reviews, as our initial results reveal. These reviews could be either written by paid reviewers (whether or not they have to use specific keywords), or they can be predefined reviews that are written by the app developers and that have to be published by the reviewers.

% ------------------------------------------------------------------------------
% ------------------------------------------------------------------------------

\section{Fake Review Characteristics (RQ2)}
\label{sec:char}

We investigate apps affected by fake reviews, reviewers providing fake reviews, and fake reviews themselves. In particular, we study the differences of fake reviews to the reviews from the official reviews dataset.

% ------------------------------------------------------------------------------

\subsection{Apps}

%
% Dataset
We identified 3,072 apps requesting fake reviews. 
As these apps could have, e.g., been entered on review exchange portals for testing purposes either or not by their developers, we only consider apps that received fake reviews to strengthen our results.
Overall, we analyzed 1,929 (62.8\%) of the identified apps.

%
% Category
\textbf{Most apps with fake reviews fall into in the category games, nearly twice as much as regular apps.} 
Table \ref{fig:appscategory} lists the 25 app categories of the Apple App Store.
It compares apps from the fake reviews and the official reviews dataset per category.
The table depicts each categories' rank, number and percentage of apps, and percentage of reviews within the datasets.
The highest rank is assigned to the category with the most apps included.

We found that more than half of the apps with fake reviews (53\%) belong to the category ``Games'', followed by the categories ``Photo \& Video'' (5.8\%), ``Education'' (4.8\%), ``Entertainment'' (4.5\%), and ``Health \& Fitness'' (4.4\%).
The categories with least apps with fake reviews are ``Stickers'' (0.05\%), ``News'' (0.1\%), ``Catalogs'' (0.16\%), ``Newsstand'' (0.16\%), and ``Books'' (0.21\%).

Between both datasets, we found a strong, positive correlation for the distribution of apps over the categories. We compared the category ranks using the Spearman rank correlation coefficient (\textit{r$_s$} = 0.74, two-tailed p-value = 0.00002). The coefficient is used to measure the rank correlation, i.e., the statistical dependence between the rankings of two variables.

To identify categories with the highest difference between both datasets, we calculated the rank difference using the following formula. 

\begin{equation}
r\textsubscript{diff}=|rank\textsubscript{official}-rank\textsubscript{fake}|
\end{equation}

The highest differences exis for the categories 
``Social Networking'' (\textit{r\textsubscript{diff}=11}), 
``Photo \& Video'' (\textit{r\textsubscript{diff}=10}), 
``Business'' (\textit{r\textsubscript{diff}=9}), and ``Shopping'' (\textit{r\textsubscript{diff}=9}).
The 3rd category ``Business'' within the app store is, for example, only ranked 12th in the fake reviews dataset. 
Vice versa, the category ``Photo \& Video'' contains more apps in the fake reviews dataset. 
\\

\begin{table}
\centering
\renewcommand{\arraystretch}{1.3}
\caption{Category ranking, number of apps, and percentage of reviews per category within the fake reviews and official reviews dataset.}
\label{fig:appscategory}
\begin{tabularx}{\columnwidth}{lrrrrrr}
\hline\noalign{\smallskip}
Category & \multicolumn{3}{l}{Fake Reviews Dataset} & \multicolumn{3}{l}{Official Reviews Dataset}      \\
\noalign{\smallskip}\cline{2-7}\noalign{\smallskip}
  & Rank   & Apps  & Reviews  & Rank    & Apps   & Reviews \\
\noalign{\smallskip}\hline\noalign{\smallskip}
\rowcolor{Gray}Books               & 21 & 4 (0.21\%) & 0.06\%      & 19 & 25069 (1.75\%) & 0.89\% \\ 
Business            & 12 & 33 (1.71\%) & 1.72\%     & 3 & 130825 (9.15\%) & 1.35\% \\
\rowcolor{Gray}Catalogs            & 23 & 3 (0.16\%) & 0.09\%      & 23 & 10951 (0.77\%) & 0.29\% \\ 
Education           & 3 & 92 (4.77\%) & 3.80\%      & 2 & 131302 (9.18\%) & 1.74\% \\
\rowcolor{Gray}Entertainment       & 4 & 87 (4.51\%) & 4.03\%      & 5 & 79504 (5.56\%) & 5.68\% \\ 
Finance             & 17 & 17 (0.88\%) & 1.11\%     & 14 & 34684 (2.43\%) & 1.66\% \\
\rowcolor{Gray}Food \& Drink       & 15 & 19 (0.98\%) & 0.65\%     & 9 & 50944 (3.56\%) & 1.34\% \\ 
Games               & 1 & 1023 (53.03\%) & 47.57\%  & 1 & 326864 (22.86\%) & 49.95\% \\
\rowcolor{Gray}Health \& Fitn.     & 5 & 85 (4.41\%) & 5.81\%      & 8 & 54410 (3.80\%) & 3.23\% \\ 
Lifestyle           & 8 & 69 (3.58\%) & 4.82\%      & 4 & 102183 (7.15\%) & 2.46\% \\
\rowcolor{Gray}Medical             & 20 & 13 (0.67\%) & 0.65\%     & 16 & 30101 (2.10\%) & 0.42\% \\ 
Music               & 11 & 36 (1.87\%) & 1.99\%     & 11 & 43874 (3.07\%) & 2.72\% \\
\rowcolor{Gray}Navigation          & 20 & 13 (0.67\%) & 1.06\%     & 20 & 21559 (1.51\%) & 0.80\% \\ 
News                & 24 & 2 (0.10\%) & 0.06\%      & 18 & 26358 (1.84\%) & 1.64\% \\
\rowcolor{Gray}Newsstand           & 23 & 3 (0.16\%) & 0.18\%      & 25 & 1021 (0.07\%) & 0.00\% \\ 
Photo \& Video      & 2 & 112 (5.81\%) & 7.66\%     & 12 & 40034 (2.80\%) & 6.54\% \\
\rowcolor{Gray}Productivity        & 10 & 42 (2.18\%) & 1.92\%     & 10 & 44191 (3.09\%) & 3.35\% \\ 
Reference           & 18 & 14 (0.73\%) & 0.63\%     & 15 & 34465 (2.41\%) & 1.50\% \\
\rowcolor{Gray}Shopping            & 13 & 25 (1.30\%) & 2.50\%     & 22 & 15253 (1.07\%) & 2.15\% \\ 
Social Netw.        & 6 & 82 (4.25\%) & 3.69\%      & 17 & 27488 (1.92\%) & 5.41\% \\
\rowcolor{Gray}Sports              & 9 & 45 (2.33\%) & 1.79\%      & 13 & 37060 (2.59\%) & 1.14\% \\ 
Stickers            & 25 & 1 (0.05\%) & 0.01\%      & 21 & 20979 (1.47\%) & 0.01\% \\
\rowcolor{Gray}Travel              & 14 & 22 (1.14\%) & 2.08\%     & 7 & 64846 (4.53\%) & 1.30\% \\ 
Utilities           & 8 & 69 (3.58\%) & 4.69\%      & 6 & 71680 (5.01\%) & 3.61\% \\
\rowcolor{Gray}Weather             & 16 & 18 (0.93\%) & 1.43\%     & 24 & 4446 (0.31\%) & 0.82\% \\
\noalign{\smallskip}\hline\noalign{\smallskip}
         &        & $\sum$ = 1,929            &             &         & $\sum$ = 1,430,091              &  \\
\noalign{\smallskip}\hline
\end{tabularx}
\end{table}

%
% Monetization
\textbf{Apps with fake reviews are on average three times less offered as paid, compared to regular apps.} 
Developers might invest a lot of money buying fake reviews.
Therefore, we analyzed the monetization of apps in the fake reviews dataset.
We focused on the app price and in-app purchases. %, and ads.

% Purchase Price
Regarding the app price, we found that 1,799 apps (93.3\%) are offered for free, while 130 apps (6.7\%) are paid. In comparison, the app store includes 1,167,377 (81.6\%) free and 262,714 (18.4\%) paid apps.
The mean price of an app is \$2.16 with a standard deviation of 2.5.
For the app store the mean price is \$4.07 with a standard deviation of 16.7.
This difference between the mean prices is statistically significant (two-sample t-test, p$<$0.001, CI=0.99).
However, the magnitude between the differences is slightly below small, found by calculating the effect size (d=-0.160) \citep{cohen1988spa}.
Of the paid apps, 62 apps (47.7\%, cf. 39.6\% in app store) are offered for \$0.99, 39 apps (30\%, cf. 18.9\% in app store) for \$1.99, and 16 apps (12.3\%, cf. 16.2\% in app store) for \$2.99. %
The remaining apps (10\%) cost between \$3.99-\$24.99. In the app store the price of the remaining apps (25.4\%) is between \$3.99-\$999.99.

%
% In-App Purchase
In-app purchases are offered by 759 fake-reviewed apps (39.4\%). 
These apps contain 3,845 in-app offers, of which 3,186 are in-app purchases. 
On average each app includes around 4.2 in-app purchases with an average price of \$10.46. 
26.33\% of the in-app purchases are offered for \$0.99, 26.9\% are in the range of \$1.99-\$2.99, 26.6\% in the range of \$3.99-\$9.99, and 20.2\% in the range of \$10.99-\$399.99.
We were unable to automatically crawl in-app purchases, therefore we cannot compare the figures from the fake reviews to the official reviews dataset.

Also, apps could be further monetized through advertisements. We were unable to study this aspect, since no publicly available data on the number of advertisement impressions and revenue generated per impression exists.

\begin{figure}
\centering
\includegraphics[width=.85\columnwidth]{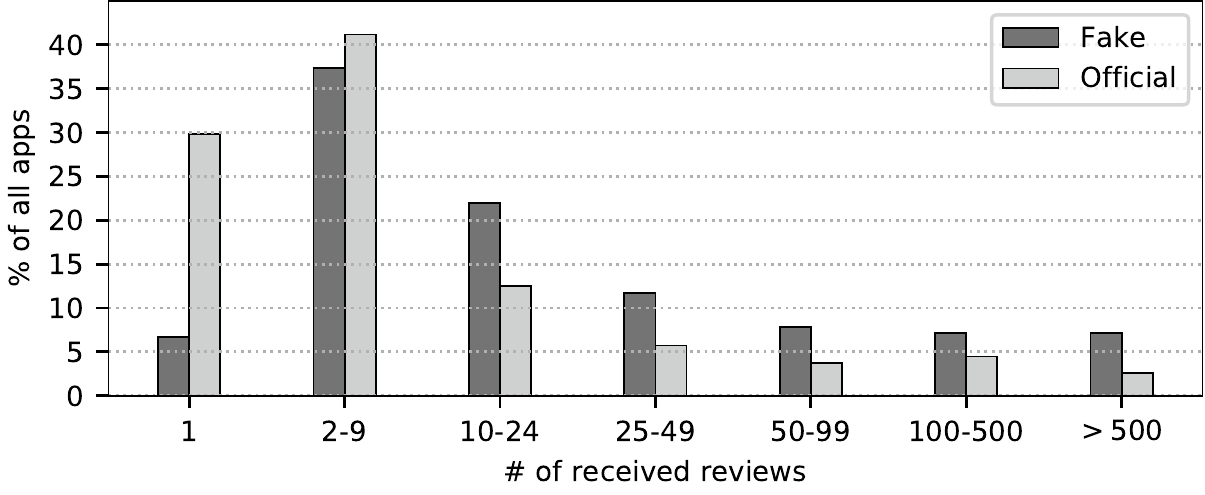}
\caption{Distribution of fake and official reviews over ranges per app.}
\label{fig:appsratio}
\end{figure}

%
% Ratio
\textbf{Most apps targeted by fake reviews have 2-9 reviews, which is the case for 42.2\% of all apps.} 
We analyzed the total number of reviews for apps affected by fake reviews and apps from the official reviews dataset.
Figure \ref{fig:appsratio} groups both types of apps into given ranges of fake and official reviews.
The result indicates that fake reviews are not necessarily limited to a small and specific group of apps, but could be distributed across the majority of apps.

\begin{figure}
\centering
\includegraphics[width=.85\columnwidth]{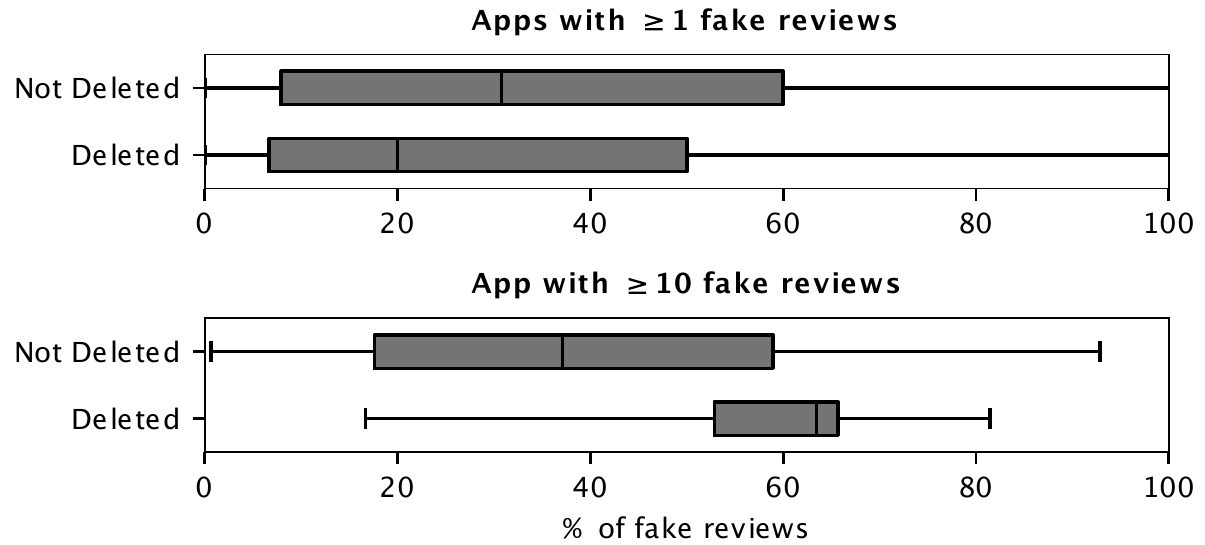}
\caption{Ratio of fake reviews for non-/deleted apps.}
\label{fig:delapps}
\end{figure}

%
% Deleted Apps
\textbf{Only less than 7\% of apps affected by fake reviews were removed from the Apple App Store.} 
We studied whether apps with a high percentage of fake reviews rather get removed from the app store, compared to apps with less fake reviews.
Therefore, we crawled the apps affected by fake reviews again after three months in June 2017.
Of the 1,929 apps, 131 (6.8\%) were no longer available on the app store. 
Most of the deleted apps (68\%) belong to the category ``Games'', 5\% to ``Entertainment'', and 5\% to ``Utilities''.
Since no justification is provided by the app store operators, there are two possible reasons for this:
Either, the apps have been removed by their own developers, or the app store operators have removed the app due to fake reviews or other compliance reasons, e.g., spam apps \citep{Seneviratne:2017:SMA:3062397.3007901}.
Figure \ref{fig:delapps} shows two plots of deleted and non-deleted apps and their percentage of fake reviews.

The upper plot considers all apps affected by fake reviews. 
We found that deleted apps received 27.5\% fake (median: 20\%) and 72.5\% official reviews.
Non-deleted apps received 38.1\% fake (median: 30.8\%) and 61.9\% official reviews.
By analyzing the median, we found that non-deleted apps receive 12 reviews (cf. 15 reviews for deleted apps) of which 2 are fake (cf. 2 reviews for deleted apps).
A $\chi^2$-test showed that being no longer available on the app store and the percentage of fake reviews are independent ($\chi^2$=2.0906, p=0.1482).

As this gives the impression that the amount of fake reviews does not impact being removed from the app store, we further analyzed apps with at least ten fake reviews only, see lower plot. 
This applies for 181 apps, of which 11 were deleted.
For these, the median of fake reviews for deleted apps is 63.5\%. 
For non-deleted apps the median is 37.1\%.
Based on medians, deleted apps receive 51 reviews of which 22 are fake. Non-deleted apps receive 49.5 reviews of which 15 are fake.
For these apps, a $\chi^2$-test showed both values are no longer independent ($\chi^2$=6.8708, p=0.008762), compared to considering all apps with at least one fake review.

% ------------------------------------------------------------------------------

\subsection{Reviewers}

%
% Average number of reviews
\textbf{Fake reviewers submit about 30 reviews on average --- 12 times more than regular reviewers.} 
We identified 721 users providing fake reviews. These fake reviewers provide 29.9 reviews per user, on average, compared to 2.5 reviews per reviewer in the official reviews dataset. 
This difference is statistically significant (two-sample t-test, p$<$0.001, CI=0.99) and the effect size is large (d=0.802).
Overall, these users provided 21,581 reviews in total for 8,429 different apps.
Surprisingly, fake reviewers do not seem to use several accounts to hide their activities.

\begin{figure}
\centering
\includegraphics[width=.85\columnwidth]{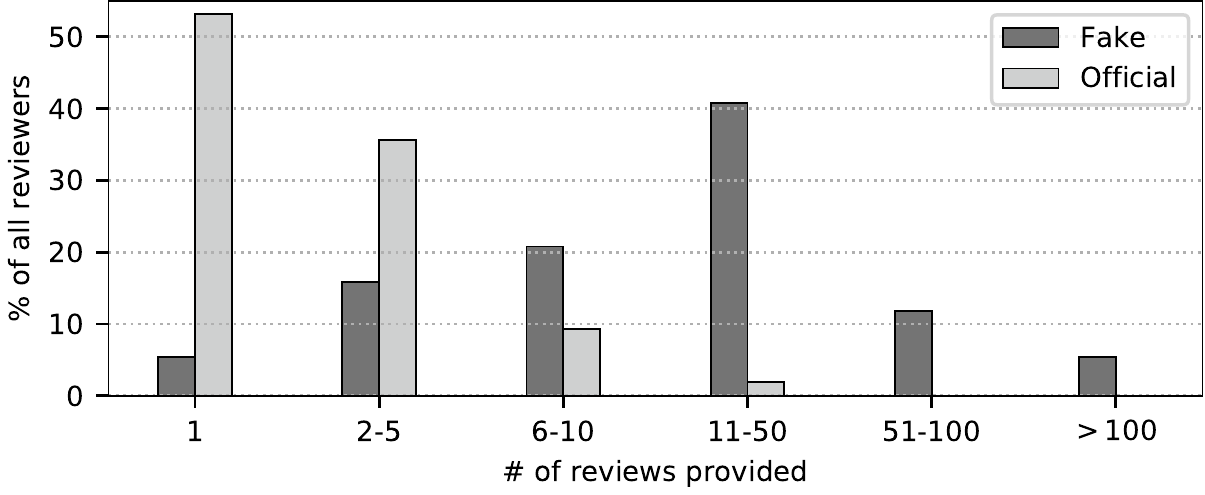}
\caption{Number of reviews provided per fake or official reviewer.}
\label{fig:reviewersgroups}
\end{figure}

%
% Total number of reviews
\textbf{More than 50\% of the reviewers in the official dataset provide only a single review.} 
The total number of reviews given per fake reviewer varies between 1 and 573. 
For reviewers within the official reviews dataset this is between 1 and 913.
Figure~\ref{fig:reviewersgroups} groups both fake and regular reviewers according to their number of submitted reviews. 

Exactly one review was given by 5.4\% of the fake reviewers, compared to 53.1\% for regular reviewers. 
2-5 reviews were provided by 15.8\% fake reviewers (cf. 35.6\%), 6-10 reviews by 20.8\% (cf. 9.3\%), 11-50 reviews by 40.8\% (cf. 1.9\%), 51-100 reviews by 11.8\% (cf. 0.02\%), and more than 100 reviews by 5.4\% (cf. 0.006\%).
The highest percentage of fake reviewers (40.8\%) is in the range of 11-50 reviews, while most regular reviewers (53.1\%) provide a single review.

%
% Review Frequency
\textbf{Fake reviewers review about 4 times more frequently than regular reviewers.} 
Fake reviewers are more active compared to others. They have a frequency of one review per 78.8 days, compared to 328.9 days for regular reviewers. 
The difference is statistically significant with 250.1 days, i.e., 417.2\% (two-sample t-test, p$<$0.001, CI=0.99).
The effect size is large (d=-0.955).

% Account usage
\textbf{The lifetime of fake reviewer accounts is nearly twice as long as regular users.} The account lifetime, i.e., the time difference between the first and last review provided, is 622.3 days for fake reviewers, compared to 331.3 days for other app store users.
The difference between fake and regular reviewers is 291 days (187.9\%) and statistically significant using the previous test (p$<$0.001, CI=0.99).
The effect size is near medium (d=0.464).
This shows that the accounts of fake reviewers remain undetected in app stores for several years.

% ------------------------------------------------------------------------------

\subsection{Reviews}

Although we found 60,431 fake reviews, in the following we only consider the 8,607 fake reviews that we identified and still exist in the Apple App Store. These reviews have not been filtered by mechanism of the app store operators and could impact app developers and users.

\begin{figure}
\centering
\includegraphics[width=.85\columnwidth]{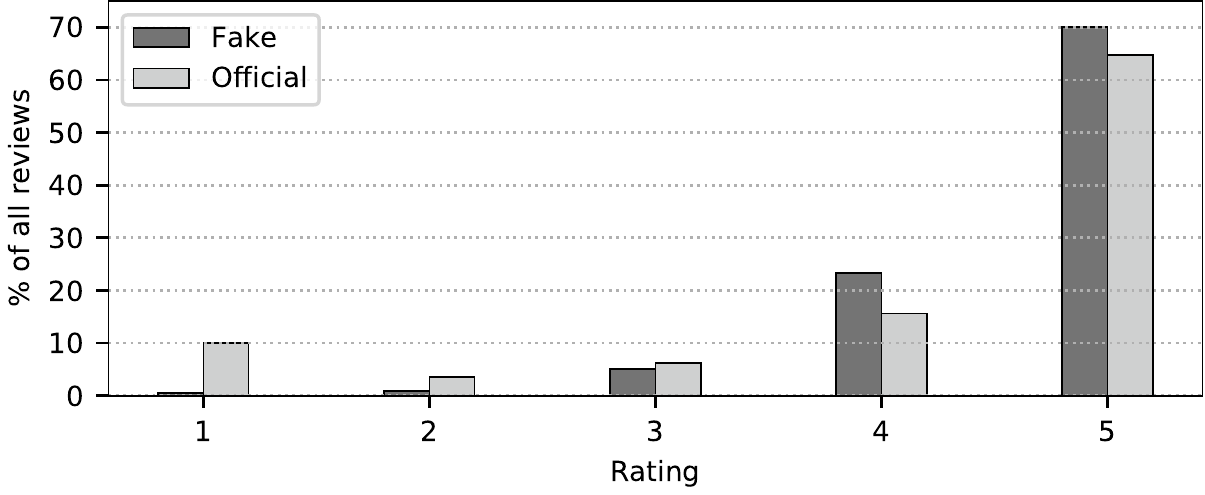}
\caption{Distribution of star ratings between official and fake reviews.}
\label{fig:ratingreviews}
\end{figure}

%
% Rating
\textbf{The distribution between ratings of fake and official reviews varies most for 1-star reviews.}
Figure \ref{fig:ratingreviews} compares the distribution of ratings for reviews from the fake and official reviews dataset.
70\% of the fake reviews are rated with 5 stars compared to 65\% for official reviews. 23\% of fake reviews are rated with 4 stars (cf. 16\%), 5\% with 3 stars (cf. 6\%), 1\% with 2 stars (cf. 4\%), and 0.6\% with 1 star (cf. 10\%). Overall, ratings are very positive in both datasets. The greatest difference between fake and official reviews can be observed by the percentage of 1-star ratings.
We have evidence that fake review providers explicitly ask their reviewers within reviewing policies to not only provide 5-stars reviews but also 4-stars and even 3-stars reviews. This might result in rather small differences between fake and official reviews regarding extremely positive ratings. Thereby, the suspicious behavior of writing, and also receiving, only 5-stars reviews should be hidden as this could possibly result in the deletion fake reviews by app store operators and in worst case the removal of the affected app from the app store (cf. Section 3.3.2).

\begin{figure}
\centering
\includegraphics[width=.9\columnwidth]{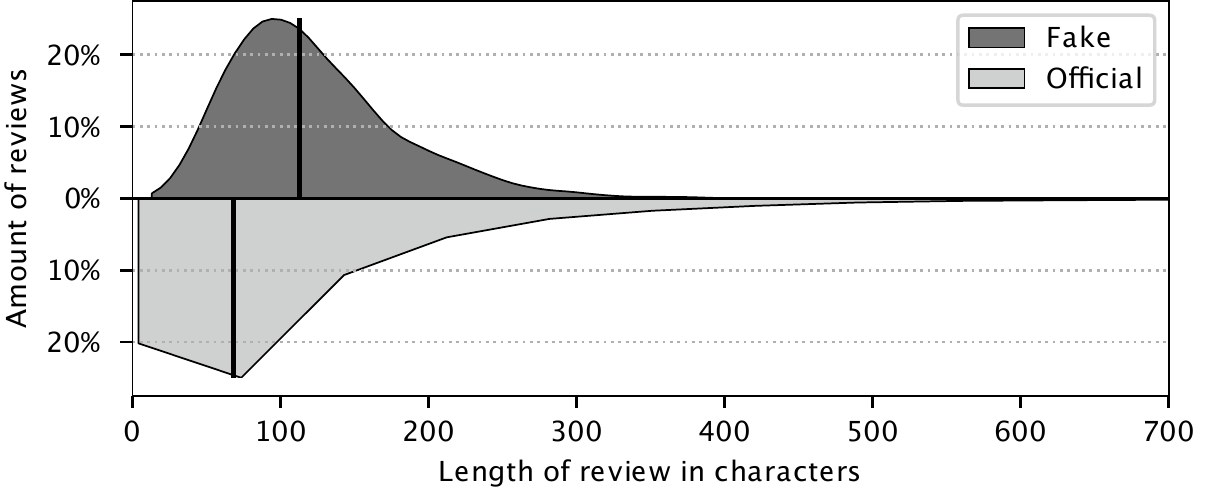}
\caption{Distribution of review length (in characters) between regular and fake reviews.}
\label{fig:reviewslength}
\end{figure}

%
% Length
\textbf{Compared to official reviews, short reviews are rather uncommon in fake reviews.}
The length of a fake review (consisting of title and body) is, on average, 121.3 characters.
Official reviews have a length of 110.8 characters, on average -- resulting in a difference of 10.5 characters.
Considering the median, fake reviews consist of 111 characters while official reviews consist of 63 characters, see Figure \ref{fig:reviewslength}. 
The difference regarding the median is 48 characters.

We further analyzed the number of words per review.
Fake reviews have, on average, 22.9 words, with a median of 21 words.
Official reviews have 21.3 words, with a median of 12 words.
Regarding the amount of average words, the difference is relatively small with 1.7 words.
Considering the median the difference is 9 words.
A typical fake review is given below.

\begin{displayquote}
Great for expense tracking $\star\star\star\star\star$\\ 
\textit{Does a great job for expense tracking. Nice interface and color scheme. Definitely recommend!}
\end{displayquote}

We found that rather short reviews, which constitute a major part of the official reviews, are uncommon for fake reviews (see example below). 

\begin{displayquote}
Fantastic $\star\star\star\star\star$\\ 
\textit{Great game, my son loves it. Lots of fun.}
\end{displayquote}

We initially assumed fake reviews to be short. However, according to the dataset fake reviews are significantly longer regarding the number of characters and words (Wilcoxon rank sum test, p$<$0.001, CI=0.99). The effect size \citep{fritz2012effect} however is near zero (r=0.001).

%
% Votes
\textbf{Fake reviews are rated more often helpful compared to official reviews.}
In app stores users can rate the helpfulness of reviews through votes. 
132 of the 8,607 fake reviews (1.5\%) received at least one vote, compared to 2.7\% for reviews in the app store.
Overall, the reviews received 270 votes.
245 are votes (90.7\%) rating the reviews as helpful and 25 votes (9.3\%) rate the reviews not helpful.
In the app store less helpful votes (67.8\%) exist. 
Both, the number of reviews with votes and the overall number of helpful votes are significantly different (two-sample t-test, p$<$0.001, CI=0.99). Also in this case, the effect size is near zero (d=-0.018).

\begin{figure}[b]
\centering
\includegraphics[width=.9\columnwidth]{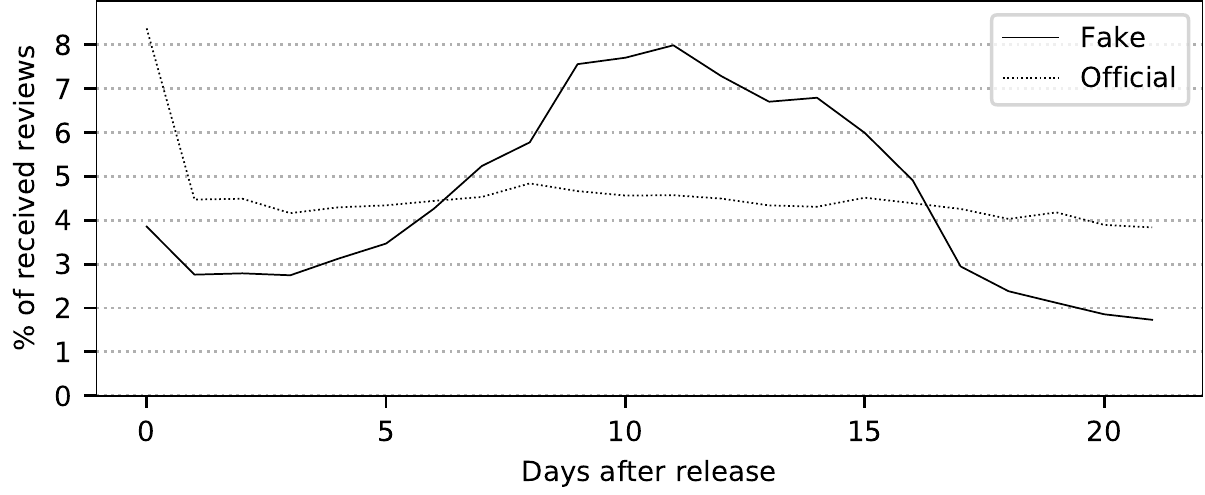}
\caption{Percentage of reviews received per day after app release (day 0 is release of app update).}
\label{fig:reviewsdate}
\end{figure}

%
% Date (e.g., Frequency, Time Difference to Release)
\textbf{After releasing updates, apps affected by fake reviews do not immediately receive more reviews.} 
We analyzed the relative reviewing frequency by summing up all reviews per day after the apps' last releases. 
Figure \ref{fig:reviewsdate} shows the percentage of received reviews per day for apps affected by fake reviews and regular apps over a time span of three weeks. 
After three weeks the amount of received reviews stabilized.
We choose only the apps' last release as we were unable to automatically crawl release dates.

For regular apps most reviews are given on the day of the app release \citep{Pagano:2013jn}. 
For apps affected by fake reviews there is only a small peak on the app release day. 
For these apps the percentage of reviews provided increases on a daily basis, until it decreases on day 11 after the app update.
One reason might be that for apps affected by fake reviews no large user basis exist that could, intrinsically motivated, provide reviews.
Developers of these apps have to buy fake reviews to promote their updates.
The distribution of request for providing fake reviews to the actual reviewers might take time.
Compared to that, regular apps with a user basis that matches the amount of reviews received, have enough users that spontaneously provide their feedback after installing the app update.

%
% Content: Most-Common Words
\textbf{Fake reviews include more positives adjectives and less negative words related to software engineering such as ``fix'' or ``crash''.} 
We analyzed the review content by comparing the 100 most-common words of fake and official reviews.
We extracted the lists of most-common words in five steps. 
First, we removed the punctuation. 
We transformed all words into lowercase writing.
Then, we tokenized the words of each review.
We removed stopwords.
Last, we counted the occurrences of each word.

\begin{figure}
\includegraphics[width=\columnwidth]{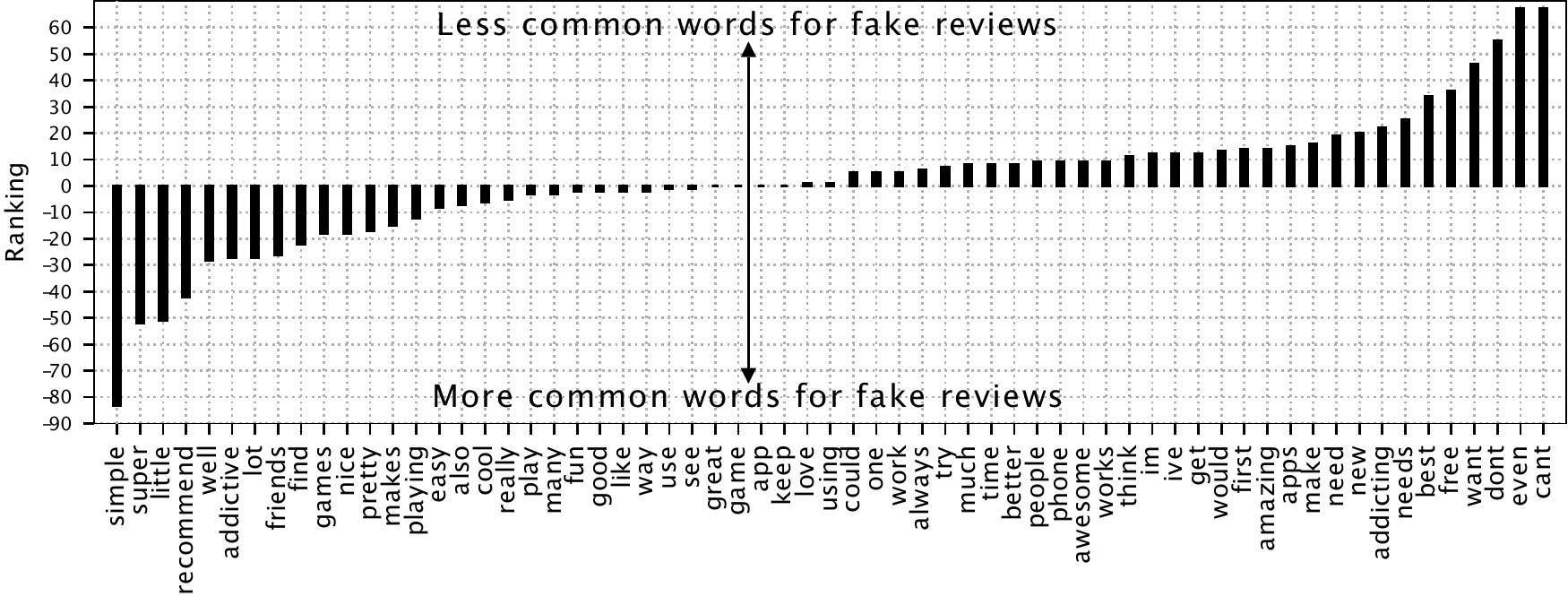}
\caption{Delta between occurrences of most-common words in official/fake reviews, a negative ranking indicates that the specific word is more common for fake reviews, a positive ranking denotes that the word is less common for fake reviews.}
\label{fig:reviewswords}
\end{figure}

Both lists have 63 words in common and 37 unique words.
We sorted the lists descending by the occurrences of words. 
Afterwards, for each word the lists have in common, we calculated the difference between their positions in the lists.
The word ``simple'' is, e.g., on position 14 for fake reviews and on position 97 for official reviews, resulting in a rank of -83.
Therefore, negative ranks denote words that are more common for fake reviews. We plotted word ranks in Figure \ref{fig:reviewswords}.

The top five words that are more common for fake reviews are ``simple", ``super", ``little", ``recommend", and ``well".
The top five words that are less common for fake reviews are ``even", ``can't", ``don't", ``want", and ``free". 

Afterwards, we analyzed the distinct words in both lists. For official reviews the distinct five most common words are (in order):  ``update'', ``ever'', ``please'', ``fix'', ``every''. Also, words possibly related to the functionality of the apps, such as ``doesn't'', ``crashes'', ``wish'', and ``bad'' are included. The five most common distinct words for fake reviews are: ``graphics'', ``useful'', ``idea'', ``ads'', and ``kids''. Also positive words, such as ``interesting'', ``perfect'', ``helpful'', ``recommended'', ``funny'', and ``learn'' are popular.

% Content: Most-Common Bi-Grams
Last, we compared the most common bi-grams for fake and official reviews. As for most common words, again 63 matches exist. We observed that bi-grams possibly pointing to bug reports, such as ``please fix'', only exist in the official reviews dataset. Bi-grams indicating feature requests, such as ``would like'' or ``wish could'', exist in both datasets. Negative bi-grams, such as ``waste time'' or ``keeps crashing'', again only exist within the official reviews dataset.

% ------------------------------------------------------------------------------
% ------------------------------------------------------------------------------

\section{Fake Review Detection (RQ3)}
\label{sec:detection}

\begin{figure}[h]
\includegraphics[width=\columnwidth]{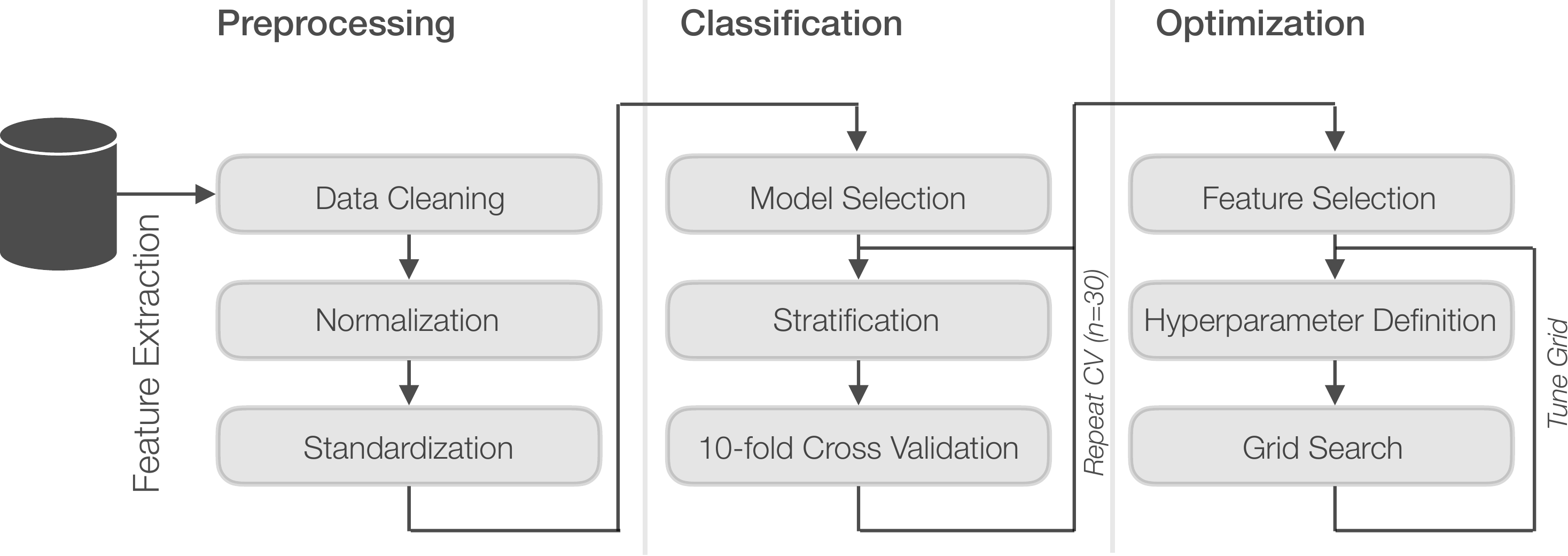}
\caption{Overview of fake review classification.}
\label{fig:classificationmethod}
\end{figure}

We build a supervised binary classifier to classify reviews as fake or not.
Figure~\ref{fig:classificationmethod} shows the three phases conducted after feature extraction. 
We begin by preprocessing the data. 
Then, we compare the results of different classification algorithms. 
We optimize the algorithms by feature selection and hyperparameter tuning. 
Last, we evaluate the importance of the classification features.
Afterwards, we conduct an in-the-wild experiment to evaluate how our classifier performs in practice, i.e., on imbalanced data.

% ------------------------------------------------------------------------------

\subsection{Feature Extraction}

We extracted a balanced truthset of 16,000 reviews. Of these reviews, 8,000 are randomly selected fake and 8,000 are randomly selected official reviews. 
Per review, the truthset includes a vector containing 15 numerical features and a label, which is either ``real'' or ``fake''.
The features were selected based on the differences we identified between fake and official reviews, and by experimentation. 
We decided \textbf{not} to use the textual review itself, e.g., in form of TF/IDF representation, for several reasons.
Mukherjee et al.~\citep{Mukherjee2013WhatYF} analyzed fake reviews published on Yelp and found that the word distribution of fake reviews does not significantly differ from official reviews. As a result, their text-based classifier only achieved an accuracy of 67.8\%.
Vice versa, the word distribution within fake reviews could also highly differ. According to the questionnaire with paid review providers, custom keywords can be included in the reviews or predefined reviews can be submitted by the fake reviewers.
Finally, when using text, training data would be required for every language, which is difficult to collect.

In contrast, Ferrara et al.~\citep{DBLP:journals/corr/FerraraVDMF14} use non-textual features related to the user, such as the account creation time or total number of followers.
By using such features their classifier to detect bots in social networks that, e.g., influence political discussions, achieved better results.
Other researchers \citep{6921650, Lee:2010:USS:1835449.1835522, feng2012distributional, Park:2016:SCM:2858036.2858389} followed this approach and were also able to improve their classification results.

We therefore focused on features that relate to the context of the fake review, i.e., the reviewer and app. 
Table~\ref{tab:featuresselected} lists all selected features. 
For the reviewer we selected four features: the total number of reviews provided, the percentage of reviews per star rating (e.g., the reviewer could have provided 70\% of all reviews with a 5-star rating and 30\% with a 1-star rating), the review frequency (i.e., the average time in seconds between all reviews provided), and the account usage (which is the lifetime of the reviewers account, i.e., the timespan between the first and the last review provided in seconds).
For the app we selected two features: the total number of reviews received for all app versions and the percentage of reviews received per star rating.
Finally feature for the review, we selected the length, i.e. the characters count.

\begin{table}[]
\centering
\renewcommand{\arraystretch}{1.3}
\caption{Features selected for the classification of fake reviews.}
\label{tab:featuresselected}
\begin{tabularx}{\columnwidth}{llXXl}
\noalign{\smallskip}\hline\noalign{\smallskip}
Category & Name                         & Type  & Null-Values & Example                       \\
\noalign{\smallskip}\hline\noalign{\smallskip}
Reviewer & \cellcolor{Gray}\# Reviews (Total)           & \cellcolor{Gray}Int   & \cellcolor{Gray}0           & \cellcolor{Gray}100                           \\
         & \% Reviews (per Star-Rating) & {[}Float{]} & 0           & {[}0.7, 0.0, 0.0, 0.0, 0.3{]} \\
         & \cellcolor{Gray}Review Frequency (in Seconds)             & \cellcolor{Gray}Int   & \cellcolor{Gray}1,734     & \cellcolor{Gray}100                           \\
         & Account Usage (in Seconds)                & Int   & 0           & 600                           \\
\noalign{\smallskip}\hline\noalign{\smallskip}
App      & \cellcolor{Gray}\# Reviews (Total)           & \cellcolor{Gray}Int   & \cellcolor{Gray}0           & \cellcolor{Gray}100                           \\
         & \# Reviews (per Star-Rating) & {[}Float{]} & 0           & {[}0.2, 0.2, 0.2, 0.2, 0.2{]} \\
\noalign{\smallskip}\hline\noalign{\smallskip}
Review   & \cellcolor{Gray}Length (in Characters)       & \cellcolor{Gray}Int   & \cellcolor{Gray}0           & \cellcolor{Gray}100                          \\
\noalign{\smallskip}\hline
\end{tabularx}
\end{table}

% ------------------------------------------------------------------------------

\subsection{Data Preprocessing}

We preprocessed the data in three steps.
We began by performing data cleaning, i.e., filling null values instead of removing affected columns.
Of the selected features only a single column includes null values, see Table~\ref{tab:featuresselected}.
The review frequency is in 1,734 cases undefined because only a single review was provided by the reviewer.
In this case, we set the frequency to lifetime of the app store, which is 9 years.

Then, we normalized the dataset so that individual samples to have unit norm.
We used the \texttt{normalize()} method with standard parameters of the \texttt{preprocessing} module provided by scikit-learn \citep{scikit-learn}.

Last, we standardized the dataset so that the individual features are standard normally distributed, i.e., gaussian with zero mean and unit variance.
This is a common requirement for many classification algorithms, such as the radial basis function (RBF) kernel of support vector machines.
If not standardizing the data, features with a much higher variance compared to others might dominate the objective function.
As a result, the classification algorithm is unable to learn from other features \citep{Weblink:SciKit:10001}.
We used the \texttt{scale()} method with standard parameters of the \texttt{preprocessing} module.

% ------------------------------------------------------------------------------

\subsection{Classification with Balanced Data}

We compare seven supervised machine learning approaches to classify reviews as fake or not.
We use the implementations provided by the scikit-learn \citep{scikit-learn} library. 
Supervised approaches need to be trained using a labeled truthset before they can be applied. 
This truthset is split into a training and testing set.
The training set is used by the algorithms to build a model on which unseen instances are classified. 
In the test phase, the classifier performs a binary classification and decides whether reviews within the test set are fake or not.

To get more reliable measures of the model quality, we apply cross validation on our truthset.
This is performed in several folds, i.e., splits of the data, called k-fold cross validation.
In this paper we perform 10 folds.
Per fold, a randomly selected amount of 1/k of the overall data is held out of the training as a test set for the evaluation. 
The final performance is the average of the scores computed in all folds.
Using cross validation, we also avoid bias which would otherwise be introduced by using only a random train/test split. 
In addition, although our truthset is balanced, we apply stratification.
Stratification ensures that each split contains a balanced amount of fake and official reviews.
We repeat the cross validation 30 times per classification algorithm with different seeds.
We use the \texttt{RepeatedStratifiedKFold} method of the \texttt{model\_selection} module.

\begin{table}[]
\centering
\renewcommand{\arraystretch}{1.3}
\caption{Classifiers' scores to detect fake reviews.}
\label{tab:classifiercompare}
\begin{tabularx}{\columnwidth}{lXXXXX}
\hline\noalign{\smallskip}
Classifier    & Accuracy & Precision & Recall & F1 & AUC/ROC \\
\noalign{\smallskip}\hline\noalign{\smallskip}
\rowcolor{Gray}RandomForestClassifier & 0.970              & 0.973             & 0.967           & 0.970       & 0.989            \\
DecisionTreeClassifier & 0.953              & 0.949             & 0.957           & 0.953       & 0.953            \\
\rowcolor{Gray}MLPClassifier          & 0.919              & 0.921             & 0.916           & 0.918       & 0.969            \\
SVC(kernel='rbf')      & 0.901              & 0.879             & 0.930           & 0.904       & 0.959            \\
\rowcolor{Gray}SVC(kernel='linear')   & 0.899              & 0.878             & 0.926           & 0.902       & 0.960            \\
LinearSVC              & 0.895              & 0.861             & 0.941           & 0.900       & 0.964            \\
\rowcolor{Gray}GaussianNB             & 0.765              & 0.731             & 0.889           & 0.755       & 0.955            \\
\noalign{\smallskip}\hline
\end{tabularx}
\end{table}

The seven classification algorithms we compare are the following:
% NaiveBayes
Naive Bayes (GaussianNB) is a popular algorithm for binary classification \citep{Bird:2009:NLP:1717171}, which is based on the Bayes theorem with strong independence assumptions between features. Compared to other classifiers it does not require a large training set.
% RandomForest
Random Forest (RF) \citep{Ho:1995:RDF:844379.844681} is an ensemble learning method for classification and other tasks. 
It can build multiple trees in randomly selected subspaces of the feature space.
% DecisionTree
Decision Tree (DT) \citep{Torgo:2010:DMR:1951702} assumes that all features have finite discrete domains and that there is a single target feature representing the classification (i.e., the tree leaves).
% SVM (linear, rbf)
Support Vector Machine (SVM) \citep{Cortes1995} represents the training data as points in space. It creates support vectors for gaps between classes in the space. The test data is classified based on which side of the gap its instances fall. The Gaussian radial basis function (rbf) is used for non-linear classification by applying the kernel trick \citep{10016858207}.
% LinearSVC
Linear support vector classification (LinearSVC) penalizes the intercept, in comparison to SVM.
% MLP
Multilayer perceptron (MLP) is an artificial neural network which consists of at least three layers of nodes. 
MLP utilizes the supervised learning technique backpropagation for training.

% Results
Table \ref{tab:classifiercompare} shows the results of the seven classification algorithms, each with default configuration.
The results include accuracy, precision, recall, F1-score, and area under the ROC curve (AUC) value. 
Among all, the random forest algorithm achieved the best scores.

% ------------------------------------------------------------------------------

\subsection{Optimization}

We optimize the classifiers by performing feature selection and hyperparameter tuning.
We optimize for \textbf{precision only}.
In comparison to fake reviews, for regular reviews we were unable to create a gold-standard dataset. 
To create a gold-standard dataset for regular non-fake reviews, all fake reviews must be identified and removed from the official reviews dataset. 
This is practically in-feasible as there is currently no measure to ensure that a review is not fake.

Hence we do not know all fake reviews, we can only report on how many of the \textit{known} fake reviews are classified as fake (precision) and not on how many of \textit{all existing} fake reviews were classified as fake (recall). Resultant measures, such as the F1-score, are reported for completeness.

To select features, we apply recursive feature elimination with cross validation.
After the classification algorithm assigned a weight to each feature, these are eliminated recursively by considering smaller sets.
Per iteration, the least important feature is removed to determine their optimal number.
We use the \texttt{RFECV} method from the \texttt{feature\_selection} module. 
The cross validation is performed as described in the previous phase.

\begin{figure}
\centering
\includegraphics[width=.8\columnwidth]{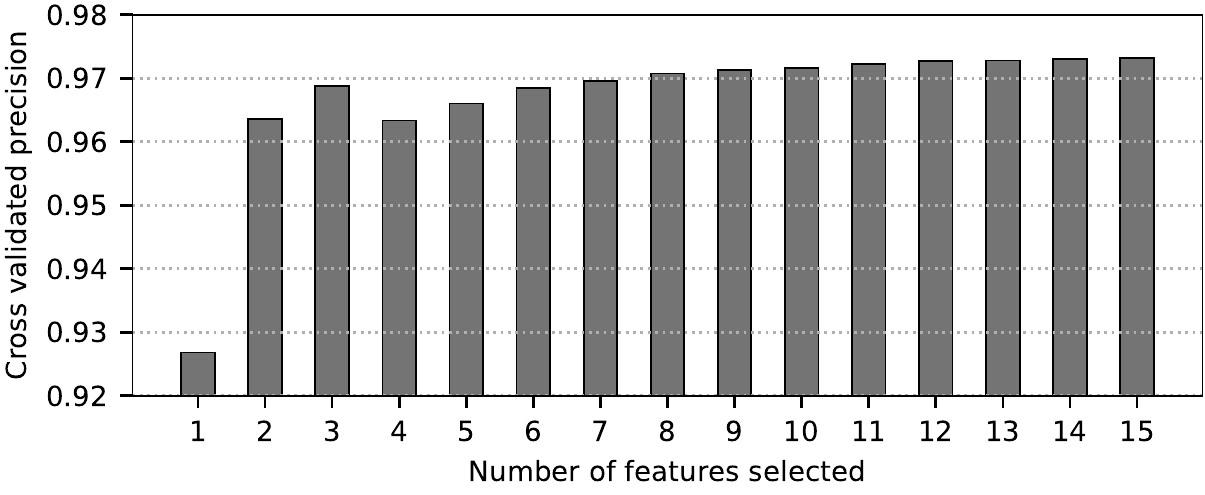}
\caption{Precision of random forest classifier by number of features selected.}
\label{fig:numfeaturesselected}
\end{figure}

We received the best result with the random forest algorithm using all features. 
Nearly similar accuracies are already possible with less features, e.g., the precision with three features is 0.969, compared to 0.973 using all features, see Figure~\ref{fig:numfeaturesselected}. The three selected features are the \textit{1) total number of reviews the app received and 2) the user provided}, as well as the \textit{3) frequency in which the user provides reviews}. 

To tune the hyperparameters, we apply the grid search method \texttt{GridSearchCV} from the \texttt{model\_selection} module.
This method performs a cross validated, exhaustive search over a predefined grid of parameters for a classification algorithm. 
After finding the optimal combination of parameters within the grid, this is further manually tuned by adding more values around the currently best.

We achieved the best result using the random forest algorithm with the parameters \textit{\{'criterion': 'gini', 'max\_depth': 30, 'max\_features': 'sqrt', 'n\_estimators': 300\}}.
The parameter \textit{criterion} measures the quality of a split. 
We used Gini impurity, which is intended for continuous attributes and faster to compute compared to Entropy. 
It is recommended to minimize the number of misclassifications.
\textit{max\_depth} defines the depth of the tree.
\textit{max\_features} sets the number of features to consider when looking for the best split.
\textit{n\_estimators} defines the number of trees in the forest.

Although performing hyperparameter tuning, the classifier's precision equal the results using the default configuration. Only measures we do not consider, the recall and F1-score, were slightly improved resulting in 98\% each.

% ------------------------------------------------------------------------------

\subsection{Feature Importance}

Last, we analyzed the relative importance of the extracted features with respect to the predictability of whether a given review is fake or not, called feature importance. 
The feature importance is calculated on how often a feature is used in the split points of a tree. 
More frequently used features are more important. 

\begin{figure}
\centering
\includegraphics[width=.8\columnwidth]{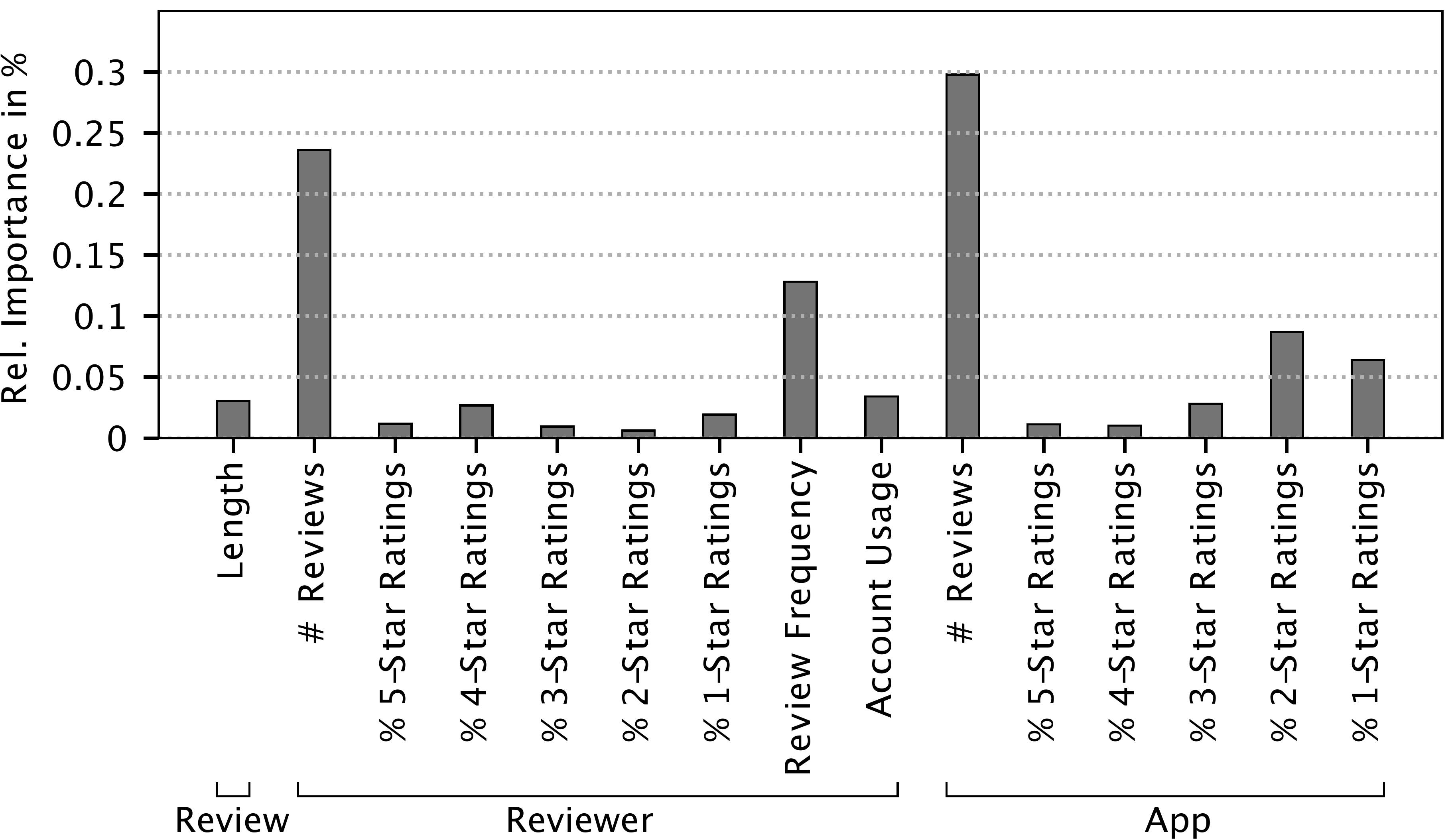}
\caption{Relative importance of extracted features to detect fake reviews.}
\label{fig:featureimportance}
\end{figure}

Figure~\ref{fig:featureimportance} shows that the three most important feature are the total number of reviews an app received (30\%), the total number of reviews a user provided (24\%), and the frequency with that a user provides reviews (13\%).

We assume the total number of reviews received by an app is the most important feature as apps with a specific amount of reviews, e.g., 2-9 reviews (cf. Section 4.1), are most often targeted by fake reviews.
The total number of reviews a user provided as well as review frequency have a high importance as fake reviewers provide much more reviews than regular reviewers, with a higher frequency.
The percentage of 1- and 2-star ratings an app received are important, as the difference between those star ratings provided are the highest when comparing apps with fake reviews and regular apps (cf. Figure \ref{fig:ratingreviews}).

% ------------------------------------------------------------------------------

\subsection{Classification with Imbalanced Data}

In practice, fake and regular reviews are imbalanced.
For app stores no reliable estimate on the distribution exists. 
Other domains, such as social media, mark 10\% to 15\% of their reviews as fake~\citep{Weblink:10019}.
The travel portal Yelp filters about 15\% of their reviews as suspicious~\citep{doi:10.1287/mnsc.2015.2304, mukherjee2013fake}.
This class imbalance can additionally be affected by numerous factors, such as the selected apps or time period.
Free apps, for example, receive more fake reviews than paid apps (cf. Section~4.1).
This reveals a skewed distribution of fake and regular reviews in app stores. 

Research found that highly imbalanced data often results into poor performing classification models~\citep{drummond2003c4, chawla2004special, saito2007large}. 
To have a more realistic setting of how our classifier can perform in practice, we conduct an in-the-wild experiment by varying the skewness of our dataset. 
We decided to vary the skewness on a logarithmic scale to depict the classification scores on finer granularities towards extremely imbalanced datasets with fake reviews as the minority class.
We keep a fixed amount of 8,000 fake reviews and create 27 datasets including $10^2-10$ (90\%) to $10^{-1}$ (0.1\%) fake reviews.
For a skew of 90\% we used 889 regular reviews.
With every change of the skewness we added additional regular reviews.
All of the about 8 million used regular reviews were randomly selected at once from the official reviews dataset, so that the classification results are comparable.

\begin{figure}
\centering
\includegraphics[width=.95\columnwidth]{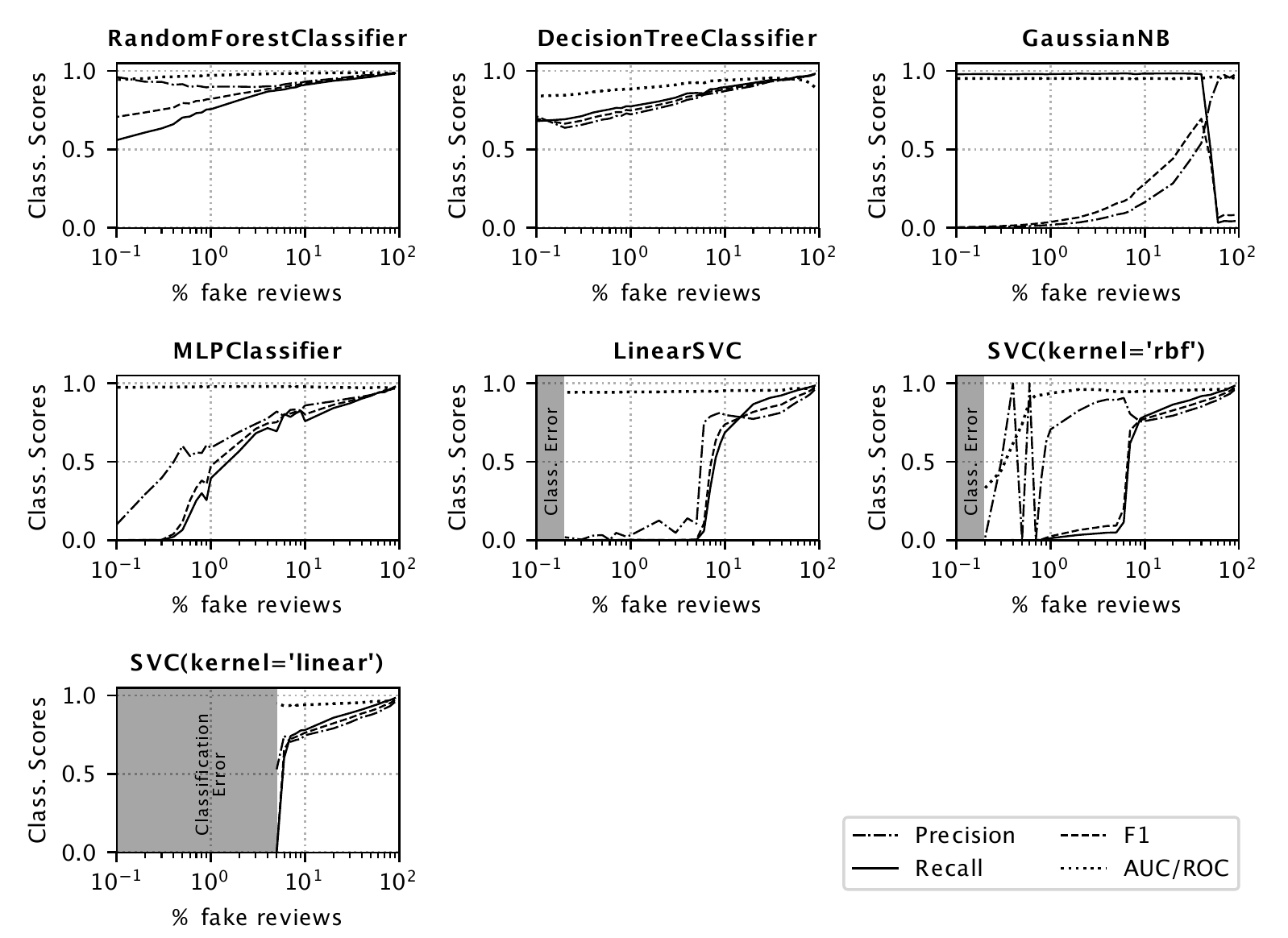}
\caption{Classification scores of machine learning algorithms on imbalanced datasets, including 90\% to 0.1\% fake reviews, plotted on a logarithmic scale.}
\label{fig:classifierskewness}
\end{figure}

Figure~\ref{fig:classifierskewness} shows an overview of the results of the seven supervised machine learning approaches studied in this work.
Per classification algorithm and skewed dataset, we report on the precision, recall, F1-score, and AUC value.

\begin{table}
\centering
\renewcommand{\arraystretch}{1.3}
\caption{Confusion matrix of a two-class problem}
\label{tab:confusionmatrix}
\begin{tabularx}{\columnwidth}{lXX}
\hline\noalign{\smallskip}
                        & Predicted as fake review  & Predicted as regular review \\
\noalign{\smallskip}\hline\noalign{\smallskip}
\rowcolor{Gray}Actual fake review      & True positive (TP)        & False negative (FN) \\
Actual regular review   & False positive (FP)       & True negative (TN) \\
\noalign{\smallskip}\hline
\end{tabularx}
\end{table}

Since identifying fake reviews is a two-class problem, the performance metrics can be derived from the confusion matrix that is generated with every classification, see Table~\ref{tab:confusionmatrix}.
We explain the reported performance metrics in the following:
%% Value: Precision
Precision ($\frac{TP}{TP + FP}$) measures the exactness, i.e., the number of correctly classified fake reviews to the overall number of reviews classified as fake.
%\begin{equation}
%    Precision = \frac{TP}{TP + FP}
%\end{equation}
%% Value: Recall
Recall ($\frac{TP}{TP + FN}$) measures the completeness, i.e., the number of correctly classified fake reviews to the overall number of fake reviews.
%\begin{equation}
%    Recall = \frac{TP}{TP + FN}
%\end{equation}
%% Value: F1
The F1-score ($\frac{recall * precision}{precision + recall}$) is the harmonic mean between precision and recall.
As improving precision and recall can be conflicting, it shows the trade-off between both.
%\begin{equation}
%    F1 = \frac{recall * precision}{precision + recall}
%\end{equation}
%% Value: AUC
The AUC value measures the area under the ROC curve.
It varies within the interval [0, 1].
The ROC curve itself depicts all possible trade-offs between TP rate, i.e., recall, and FP rate ($\frac{FP}{TN + FP}$).
%TP and FP rate can be understood as the benefits and costs of classification with respect to data distributions.
A better classifier produces an ROC curve closer to the top-left corner and therefore a higher AUC value~\citep{8246564}.
%\begin{equation}
%    FPR = \frac{FP}{TN + FP}, TPR = \frac{TP}{TP + FN} = recall
%\end{equation}

% Exclude scores (class imbalance)
As we are focusing on skewed datasets within our in-the-wild experiment, we need to select performance metrics that are insensitive to class imbalance.
This applies for all measures that use values from only one row of the confusion matrix~\citep{8246564}.
Precision and F1-score are sensitive to class imbalance and biased towards the majority class. Therefore, these metrics are inappropriate for our evaluation.
Recall and AUC/ROC value are insensitive to class distribution, we use both measures to compare the performance of the classification algorithms within our in-the-wild experiment.

% Include classsifiers
%% RandomForestClassifier
%% DecisionTreeClassifier
%% MLPClassifier
In the further evaluation, we include all classifiers that achieve a recall and AUC value higher than 0.5 for datasets including 90\% to 1\% fake reviews.
This applies for the random forest (RF), decision tree (DT), and MLP algorithm.
The remaining algorithms are excluded from the evaluation.
%% GaussianNB
The recall of the gaussian naive bayes (GaussianNB) algorithm decreases to nearly 0 for datasets with fake reviews as the majority class. 
Also, its precision is extremely low for datasets including less than 10\% fake reviews.
%% LinearSVC
%% SVC(kernel=linear)
%% SVC(kernel=rbf)
Similarly, the recall of the SVC algorithms decrease to nearly 0 for datasets with less than 10\% fake reviews. 
In addition, the SVC(kernel='linear') implementation of sklearn return errors for skews below 5\%, so do the remaining two SVC algorithms for skews of 0.1\%.

\subsubsection{Classification Results with Imbalanced Data}

\begin{figure}
\centering
\includegraphics[width=\columnwidth]{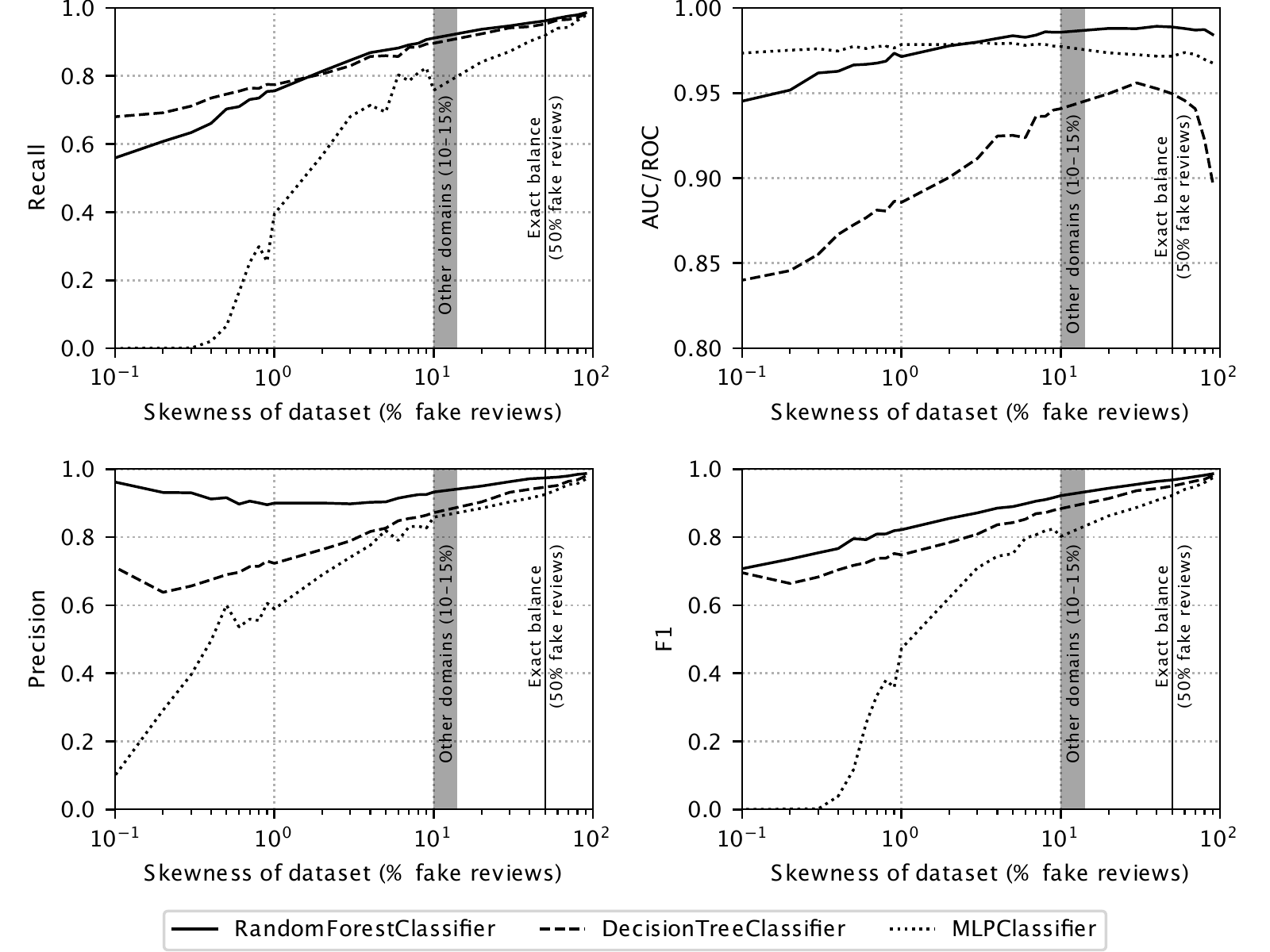}
\caption{Classification scores of appropriate machine learning algorithms for datasets with class imbalance, i.e., including 90\% to 0.1\% fake reviews, plotted on a logarithmic scale.}
\label{fig:classifierskewnessrel}
\end{figure}

Figure~\ref{fig:classifierskewnessrel} shows the performance measures of all three classification algorithms that remain within our evaluation.
We chose to depict each measure as a single plot to more easily compare the algorithms.
We report the precision and F1-score for reasons of completeness, although these are inappropriate measures for imbalanced data.
Within the graphs, we highlight in gray the interval in which fake reviews typically occur in other domains.
Further, we mark where the classes are equally distributed, i.e., there exist 50\% fake reviews.
At this point the algorithms perform well with all measures above 0.9 (cf. Table~\ref{tab:scoresrel}).

\begin{table}[]
\centering
\renewcommand{\arraystretch}{1.3}
\caption{Classification scores on imbalanced datasets with skews of 90\% to 0.1\% fake reviews (DT: DecisionTreeClassifier, MLP: MLPClassifier, RF: RandomForestClassifier).}
\label{tab:scoresrel}
\begin{tabularx}{\columnwidth}{rXXXXXXXXXXXX}
\hline\noalign{\smallskip}

Skew & \multicolumn{3}{l}{Recall} & \multicolumn{3}{l}{AUC/ROC} & \multicolumn{3}{l}{Precision} & \multicolumn{3}{l}{F1-score} \\ 
\noalign{\smallskip}\cline{2-13}\noalign{\smallskip}
     & DT    & MLP   & RF    & DT    & MLP   & RF    & DT    & MLP   & RF    & DT    & MLP   & RF      \\ 

\noalign{\smallskip}\hline\noalign{\smallskip}

\rowcolor{Gray}90.0   & 0.982 & 0.978 & 0.986 & 0.897 & 0.968 & 0.984 & 0.979 & 0.970 & 0.987 & 0.981 & 0.974 & 0.986 \\
80.0   & 0.972 & 0.963 & 0.980 & 0.923 & 0.970 & 0.987 & 0.968 & 0.958 & 0.984 & 0.970 & 0.960 & 0.982 \\
\rowcolor{Gray}70.0   & 0.966 & 0.943 & 0.976 & 0.941 & 0.973 & 0.987 & 0.964 & 0.954 & 0.980 & 0.965 & 0.949 & 0.978 \\
60.0   & 0.964 & 0.940 & 0.970 & 0.946 & 0.974 & 0.988 & 0.952 & 0.940 & 0.976 & 0.958 & 0.940 & 0.973 \\
\rowcolor{Gray}50.0   & 0.953 & 0.920 & 0.962 & 0.950 & 0.972 & 0.989 & 0.947 & 0.925 & 0.974 & 0.950 & 0.922 & 0.968 \\
40.0   & 0.945 & 0.902 & 0.956 & 0.953 & 0.972 & 0.989 & 0.941 & 0.914 & 0.972 & 0.943 & 0.908 & 0.964 \\
\rowcolor{Gray}30.0   & 0.942 & 0.872 & 0.947 & 0.956 & 0.973 & 0.988 & 0.932 & 0.903 & 0.963 & 0.937 & 0.887 & 0.955 \\
20.0   & 0.924 & 0.842 & 0.937 & 0.950 & 0.974 & 0.988 & 0.903 & 0.885 & 0.950 & 0.914 & 0.863 & 0.944 \\
\rowcolor{Gray}10.0   & 0.896 & 0.758 & 0.912 & 0.941 & 0.977 & 0.986 & 0.872 & 0.859 & 0.933 & 0.884 & 0.802 & 0.922 \\
9.0    & 0.894 & 0.824 & 0.907 & 0.940 & 0.978 & 0.986 & 0.865 & 0.826 & 0.925 & 0.879 & 0.824 & 0.916 \\
\rowcolor{Gray}8.0    & 0.885 & 0.809 & 0.896 & 0.936 & 0.979 & 0.986 & 0.859 & 0.832 & 0.925 & 0.872 & 0.819 & 0.910 \\
7.0    & 0.884 & 0.785 & 0.892 & 0.936 & 0.979 & 0.984 & 0.855 & 0.830 & 0.920 & 0.869 & 0.807 & 0.906 \\
\rowcolor{Gray}6.0    & 0.857 & 0.804 & 0.882 & 0.924 & 0.978 & 0.983 & 0.848 & 0.790 & 0.915 & 0.853 & 0.797 & 0.898 \\
5.0    & 0.860 & 0.694 & 0.876 & 0.925 & 0.979 & 0.984 & 0.826 & 0.820 & 0.904 & 0.843 & 0.752 & 0.890 \\
\rowcolor{Gray}4.0    & 0.857 & 0.714 & 0.869 & 0.925 & 0.979 & 0.982 & 0.816 & 0.776 & 0.902 & 0.836 & 0.744 & 0.885 \\
3.0    & 0.830 & 0.681 & 0.846 & 0.912 & 0.980 & 0.980 & 0.789 & 0.741 & 0.898 & 0.809 & 0.710 & 0.871 \\
\rowcolor{Gray}2.0    & 0.806 & 0.567 & 0.814 & 0.901 & 0.978 & 0.978 & 0.764 & 0.690 & 0.900 & 0.784 & 0.622 & 0.855 \\
1.0    & 0.775 & 0.395 & 0.756 & 0.886 & 0.978 & 0.971 & 0.723 & 0.589 & 0.900 & 0.748 & 0.473 & 0.822 \\
\rowcolor{Gray}0.9  & 0.776 & 0.255 & 0.755 & 0.886 & 0.976 & 0.973 & 0.730 & 0.606 & 0.895 & 0.752 & 0.359 & 0.819 \\
0.8  & 0.764 & 0.300 & 0.735 & 0.881 & 0.978 & 0.969 & 0.714 & 0.555 & 0.901 & 0.738 & 0.380 & 0.810 \\
\rowcolor{Gray}0.7  & 0.765 & 0.252 & 0.731 & 0.881 & 0.977 & 0.968 & 0.714 & 0.559 & 0.906 & 0.738 & 0.334 & 0.809 \\
0.6  & 0.756 & 0.166 & 0.710 & 0.877 & 0.976 & 0.967 & 0.697 & 0.536 & 0.897 & 0.725 & 0.253 & 0.793 \\
\rowcolor{Gray}0.5  & 0.747 & 0.065 & 0.703 & 0.873 & 0.977 & 0.966 & 0.690 & 0.600 & 0.916 & 0.717 & 0.117 & 0.796 \\
0.4  & 0.735 & 0.021 & 0.661 & 0.867 & 0.975 & 0.963 & 0.675 & 0.496 & 0.912 & 0.704 & 0.038 & 0.766 \\
\rowcolor{Gray}0.3  & 0.712 & 0.001 & 0.634 & 0.855 & 0.976 & 0.962 & 0.656 & 0.396 & 0.930 & 0.683 & 0.002 & 0.754 \\
0.2  & 0.692 & 0.001 & 0.608 & 0.846 & 0.975 & 0.952 & 0.638 & 0.292 & 0.931 & 0.664 & 0.001 & 0.736 \\
\rowcolor{Gray}0.1  & 0.680 & 0.000 & 0.560 & 0.840 & 0.973 & 0.945 & 0.712 & 0.100 & 0.962 & 0.696 & 0.000 & 0.707 \\ 

\noalign{\smallskip}\hline
\end{tabularx}
\end{table}

% Imbalanced data
Unfortunately, in practice there can exist imbalances towards fake or regular reviews being the majority class.
From our results and research in other domains it is more likely that the bias is towards regular reviews.
For this reason we choose more detailed results towards fake reviews being the minority class.

%% FR majority
However, when fake reviews become the majority class (towards the right of the 50\% mark) the recall improves for all algorithms, by up to 6.4\% for the MLP algorithm. The best result is achieved by the RF algorithm with 0.986 recall. 
The AUC value decreases in all cases, for the RF and MLP algorithms the value slightly decreases by up to 0.5\%.
The value of the DT algorithm decreases more stronly by 5.5\%.
The best AUC value is achieved by the random forest algorithm with 0.984.

%% FR minority
When fake reviews become the minority class most performance measures decrease.
%%% 10^1 = 10%
With an amount of \textbf{10\% fake reviews} ($10^1$), as reported to be typical for other domains~\citep{Weblink:10019,doi:10.1287/mnsc.2015.2304, mukherjee2013fake}, the recall of the RF and DT algorithm are nearly identical (RF: 0.912, DT: 0.896).
Compared to the result of the balanced dataset, the recall decreased by 5.3\% for the RF algorithm and by 5.9\% for the DT algorithm.
The recall of the MLP algorithm is significantly less with 0.758 (-17.6\%).
The AUC value is the highest for the RF algorithm (0.986, +0.3\%), followed by the MLP (0.977, +0.6\%) and DLT (0.941, -0.9\%) algorithms.

%%% 10^0 = 1%
For \textbf{1\% fake reviews} ($10^0$), the recall of the RF and DT algorithms is still nearly identical (RF: 0.756, -21.4\%; DT: 0.775, -18.7\%), followed by the MLP algorithm (0.395, -57.1\%).
The AUC value is the highest for the MLP algorithm (0.978, +0.7\%), followed by the RF (0.971, -1.8\%) and DT (0.886, -6.7\%) algorithms.

%%% 10^-1 = 0.1%
For an amount of \textbf{0.1\% fake reviews} ($10^{-1}$) the recall is the highest for the DT algorithm (0.680, -28.6\%), followed by the RF algorithm (0.560, -41.8\%). The recall of the MLP algorithm dropped to 0 at about 0.3\% fake reviews and below within the dataset.
The AUC value is the highest for the MLP algorithm (0.973, +0.2\%), followed by the RF algorithm (0.945, -4.4\%). 
Last, the AUC value of the DT algorithm significantly decreased to 0.840 (-11.4\%).

% Overall
Comparing all three algorithms using their recall and AUC/ROC value, \textbf{the random forest algorithm performs best for imbalanced datasets}. 
Although, the decision tree algorithm achieves a better recall when the dataset is extremely skewed (less than 1\% fake reviews), its AUC/ROC value is significantly lower for all datasets.
Similar, the MLP algorithm achieves better AUC/ROC values for datasets with less than 1\% fake reviews. However, the recall of the MLP algorithm drops to 0 for extremely skewed datasets.
With skews common for other domains, the random forest algorithm performs best with a recall of 0.912 and AUC/ROC value of 0.986.

% ------------------------------------------------------------------------------
% ------------------------------------------------------------------------------

\section{Discussion}
\label{sec:discussion}

We discuss implications of fake reviews on software engineering, and from the perspective of app users and store operators. 
Then, we discuss the results' validity.

\subsection{Implications}

In  modern app stores, developers are for the first time able to publicly retrieve customers' and users' opinions about their software and to compare its popularity in form of rank or number of downloads.
Although app reviews provide a rich source of information, they may not be fully reliable, as customers may leave reviews that do not reflect their true impressions \citep{FINKELSTEIN2017119}. 
Our work shed light on one of these cases: fake reviews.

Generally, fake reviews, i.e., paid, incentivized reviews (which can provided either directly or via fake review providers) are prohibited by official app store reviewing policies. 
The main reason is to \textbf{preserve the integrity of app stores}~\citep{Weblink:10005, Weblink:10020}.
Users that do not trust app stores and their reviews will most likely refrain from providing app reviews themselves.
This would harm one of the most important advantages of app stores: collecting \textit{real, spontaneous feedback} on software in a channel used by both developers and users.

We applied our fake review classifier to the full official Apple App Store dataset. As a results, 22,207,782 (35,5\%) of all 62,617,037 reviews were classified as fake.
This number seems very high at first and can only be used as a first indication. Further studies need to be carried out to give a precise approximation of the amount of fake reviews in official app stores. 
Still, multiple indices indicates a non-trivial amount of fake reviews in app stores. We identified about 60,000 reviews from only a single provider.
Overall, we identified 43 providers while much more might exists or have existed before.
If every provider would provide the same number of reviews, the amount would sum up to 2.58 million fake reviews.
We hypothesize that the majority of fake reviews is written by persons, who get directly asked by developers. 
Although not generalizable, we repeatedly observed this phenomena in our professional app development settings. 
When apps are developed privately, friends were asked to provide fake reviews.
When programmed in a commercial environment, either employees of the developing company  or of the ordering  company (cf. Bell \citep{Weblink:10007}) are asked to provide fake reviews. 
Given that 1.4 million apps exists within the dataset the number of fake reviews does no longer seem unattainably high.
Such amount of fake reviews are also presumed in other domains. Streitfeld \citep{Weblink:10015} report that every fifth review submitted to Yelp is detected as dubious by internal filters.

\textbf{App users} might, by using positive or negative fake reviews, get mislead to either downloading an app or not. As shown by Ott et al. \citep{Ott:2011:FDO:2002472.2002512} fake reviews sound authentic and are hard to detect by humans. In an experiment, humans at most scored an accuracy of 61\% identifying fake reviews, even as the word distribution of the used fake reviews differed from regular reviews. We think that this also applies for apps. Users and developers might not be able to identify fake reviews only based on their text.

Measures and tools should enable users (and developers) to identify fake reviews and affected  apps. Such tools already exist for products sold on Amazon, e.g., Fakespot \citep{Weblink:10014}. Users enter the name of a product to determine if its reviews are trustworthy. Fakespot also takes a step towards analyzing fake reviews in the app store.  The features used to classify reviews as fake also related to the review context, such as if a large number of positive reviews is provided within a short period of time. However, the selected criteria to classify reviews as fake are not completely transparent nor empirically validated. Also, we assume that no gold-standard dataset has been used for fake reviews. For the Instagram app the site, e.g., classifies 50\% (about 600,000) of the reviews as fake, which raises accuracy concerns for this approach.

The \textbf{research area App Store Analysis} covers work to mine apps and their reviews and extract relevant information for software and requirements practitioners, e.g., to get inspirations about what should be developed and to guide the development process \citep{Harman:2012gw, Pagano:2013:UIS:2486788.2486920, Pagano:2013jn, 7886888, Chen:2014:AMI:2568225.2568263, 6912257, 6624001, Khalid:2013:IUC:2486788.2487044, 6606604}.
Martin et al. provide a comprehensive literature review of this area \citep{Martin:2016fe}. 
The majority of papers identified in their study (127 of 187 papers, 68\%) analyze non-technical information, such as app reviews.

Review Analysis itself is one of the largest sub-fields of app store analysis, which receives a significant and increasing number of publications each year. 
Work in this sub-field started by analyzing the content of app reviews (2012 -- 2013), afterwards focusing on adding additional features such as sentiments (2013 -- 2014). 
Then, app reviews were summarized to extract app requirements. 
Although, information extracted from app reviews is getting increasingly integrated into the requirements engineering processes \citep{7332475, Maalej:2015:Software}, \textbf{none of the papers discusses the impact of fake reviews}. 
 In the following, we discuss the potential impact of fake reviews on software engineering along with the main review analysis topics according to Martin et al. \citep{Martin:2016fe}. 

\begin{figure}
\centering
\includegraphics[width=.85\columnwidth]{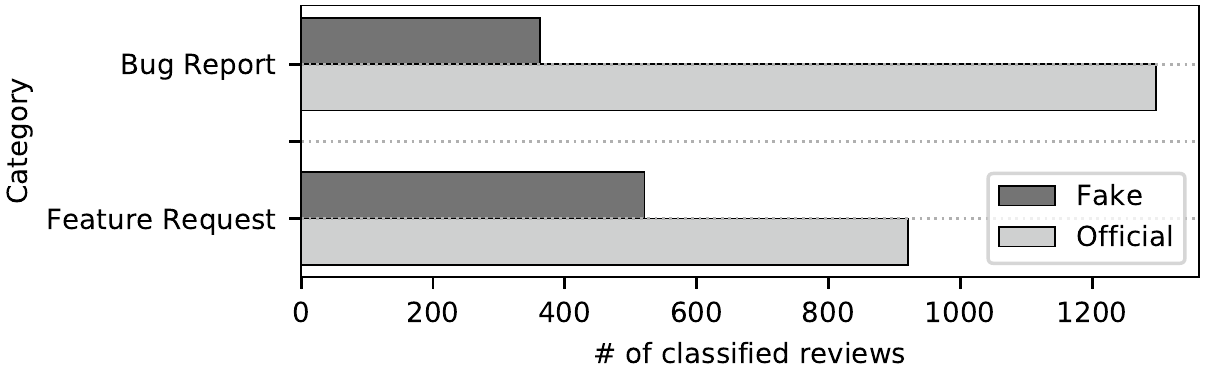}
\caption{Reviews within truthset classified as bug report and/or feature request.}
\label{fig:truthsetclassification}
\end{figure}

% 9.3 Requirements Engineering
Fake reviews, similar to official reviews, include \textbf{requirements-related information, such as feature requests.}
Oh et al. automatically categorize app reviews into bug reports and non-/functional requests to produce a digest for developers including the most informative reviews \citep{Oh:2013:FDI:2468356.2468681}. 
Additional work focuses on extracting requirements-related information from app reviews \citep{10.1007/978-3-319-05452-0_4, Iacob:2013:YCS:2578048.2578086, 7332474}. 
Iacob and Harrison found that 23.3\% of the app reviews include feature requests \citep{6624001}.
We applied the classifier of Maalej et al. \citep{Maalej:2016:Automatic} to extract bug reports and feature request from the 8,000 official and 8,000 fake reviews included in our truthset (see
Figure \ref{fig:truthsetclassification}). 
While fake and regular reviews are imbalanced within the overall app store dataset, when applying review analysis approaches on a subset of reviews the distribution is unknown. For this reason, we did not set a specific distribution of fake and regular reviews.
Within the official reviews, we identified  1,297 bug reports and 921 feature requests, while we found that the fake reviews contain 362 bug reports and 521 feature requests.
We include an example feature request within the fake review dataset  below. 

\begin{displayquote}
Nice UI $\star\star\star\star$

\textit{Very clean and beautiful UI. I like the goal setting and the reminders. I would like to see some animation when scrolling the weekly progress bars.}
\end{displayquote}

We assume that most fake reviewers did not use the reviewed app before and are unfamiliar with it.
In addition, review policies ask fake reviewers to explicitly talk about features rather than providing praise only. 
For this reasons, it is unclear if those feature requests are really relevant or only thought up to make the review sound more authentic. 
Fake reviews might thus impact the results of existing classifiers. 
When not removing or at least flagging fake reviews with  information relevant for developers, wrong assumptions for the future decisions might be drawn.

% 9.2 Content
Researchers showed that nearly \textbf{half of the feature requests included in app reviews are implemented.} 
Hoon et al. highlight that user expectations are changing rapidly, as observable through app reviews. Developers must keep up with the demand to stay competitive \citep{hoon2013analysis}.
Palomba et al. studied the reviews of 100 open-source apps. By linking reviews to code changes the authors showed that 49\% of the changes requested were implemented in app updates \citep{7332475}.
These results show that a significant amount of changes proposed by users are integrated into software.
Some of these changes might be inspired and prioritized based on fake reviews. 

% 9.5 Summarisation
Recent approaches summarize and extract requirements-related information from \textbf{app reviews of related or competing apps.}
Fu et al. present WisCom to analyze app reviews per app/market level, e.g., to get an overview of competing apps \citep{fu2013people}.
Gao et al. present AR-Tracker to summarize app reviews to real issues and prioritize them by their frequency and importance \citep{gao2015ar}.
Nayebi et al. mine app reviews and tweets of similar apps within a specific domain \citep{8049145, 7961668}.
These approaches monitor and extract information from app reviews of related or competing apps. 
While developers will probably know when their apps receive fake reviews, i.e., when they bought those instead of being affected by negative fake reviews bought by competitors, developers cannot be fully sure if competing apps receive fake reviews. 
Wrong conclusions can also be drawn from fake reviews including the honest opinion of reviewers. Fake reviewers can just copy and modify a regular review they honestly agree with and resubmit those on review exchange portals. This way the frequency (and hence the priority) of, e.g., a feature request, might be fake (i.e. incentivized) and thus biased.
Table~\ref{tab:guidelines} highlights that only three review exchange portals forbid fake reviewers to copy and modify existing reviews.
Using those reviews as input for summarization approaches, ``wrong'' features could emerge as a result.

\textbf{App store operators} try to prevent fake reviews by providing review policies. The Apple App Store policy \citep{Weblink:10005} states that apps will be removed and that the developers may be expelled from the app store's developer program \textit{"if we find that you have attempted to manipulate reviews, inflate your chart rankings with paid, incentivized, filtered, or fake feedback"}. This is the case when app developers buy fake reviews. However, the concrete actions taken to identify fake reviews are non-transparent. We also noticed that larger, popular apps try to prevent negative feedback from being submitted to the app store. These apps ask users for submitting feedback within the app in form of star ratings. The rating is not directly forwarded to the app store. In case of a one star rating, a mail form appears asking the user to submit the review directly to the app developer instead of forwarding it to the app store, where the review is publicly visible. Such actions might also manipulate the app ratings and reviews as well.

We think that researchers should carefully sample apps and perform data cleanings before studying and mining app reviews. 
For example, apps with an unusual distribution of ratings might be affected by fake reviews. 
Similar, reviews of users with an amount or frequency above average might have to be removed or considered separately during data cleaning. 
Otherwise, wrong assumptions for the future development of an app could be drawn. 

Collecting and analyzing \textbf{context and usage information} can help substantiate decisions and check the quality of reviews   \citep{Maalej:2015:Software}.
App developers can, e.g., utilize the number of users or the average time users spend with a specific feature to decide which parts of their apps to improve and which suggestions should be taken into consideration in the next release.
App store operators can take measures, such as the number of times a user opened an app or the daily app usage time to decide the trustworthiness and weight of reviews within an app's overall rating.
Instead of limiting the amount of users who can participate in the reviewing process, one might think about to weight or consider incentivized reviews differently and in a transparent manner.
Since the overall app store ecosystem is designed that more positive reviews lead to more downloads and thus increase the app's success \citep{Harman:2012gw}, developers will likely continue to ask ``friends'' to rate their apps.
Even if incentivized and not independent, such reviews can also include useful information.
Instead of excluding the reviews and their reviewers, a possible alternative might be to highlight these review with badges (e.g., friend, expert, or crowd testers).

% ------------------------------------------------------------------------------

\subsection{Results Validity}

The results of our study might have limitations and should be considered within the study context.

The process of extracting actual fake reviews was challenging. However, we did not try to generate fake reviews using a crowdsourced approach, as we wanted to only rely on fake reviews that have been published to and still remain unidentified within the app store by its users and operators. For this reason, we decided only using a subset of our about 60,000 collected fake reviews.

Nearly all reviews were extracted from a single provider. This provider is a review exchange portal (see Table~\ref{tab:fakereviewdataset}, REP3).
On these portals reviewers could submit their honest opinion.
However, \textbf{we consider the collected reviews as fake for the following reasons:}

First, app store operators strongly require that app reviews must be 1) written by \textit{real users} of the app and 2) cannot be \textit{incentivized}. Both conditions are not given in review exchange portals. For this reason, these reviews are fake according to the definition and agreement or app store providers. Even if reviewers are allowed to submit their honest opinion according to the review policy of this exchange portal, \textit{rewarded, incentivized, or non-spontaneous reviews} are prohibited by the official Google and Apple App Store Review Guidelines~\citep{Weblink:10005, Weblink:10020}.

Second, review exchange portals provide predefined ratings and review messages. 
These ratings do not necessarily correspond to the opinion of the reviewers.
The providers' review policies ensure that reviewers that post their honest opinion are not being rewarded and are excluded from reviewing portals.
The review policy of REP3 does not include a general rating, e.g., 3-stars or above (cf. Table~\ref{tab:guidelines}).
The provider uses individual policies per app (cf. Figure~\ref{fig:reviewrequest}).
We could not extract historical data to say if all individual policies included a predefined rating. 
However, for active review requests at the time of data collection, predefined ratings were included in all cases.

Third, paid review providers and review exchange portals share reviewers.
As identified, paid review providers cross-post their review requests on review exchange portals (cf. Figure~\ref{fig:reviewingstrategy}).
This also applies for REP3.
Paid review providers would not cross-post their review requests on these portals, if the app ratings would not change as desired by app developers.

Fourth, per app we compared the collected fake reviews to each other. 
We searched for apps that received fake reviews with a rating of 1-2 stars as well as fake reviews with a rating of 4-5 stars.
These reviews are most likely written by reviewers that posted their honest opinion about an app, either positive or negative.
This applies for only 32 of the 1,890 (1.69\%) collected apps.
For these apps, 41 1-star and 2-star reviews out of 8,607 reviews (0.48\%) were provided.

Another limitation is that although we filtered fake reviews for reviews in English language only and targeted the US storefront of the Apple App Store, reviews in English language could have been submitted to other storefronts.
This could be a possible reason why we were only able to identify 8,607 of the initially collected 60,431 fake reviews within the app store, i.e., the official reviews dataset.
In this case, the moderation of reviews by app store operators is less strict than observed.

Further, review exchange portals are also used by app developers, since the reward of providing a fake review is a credit which can be redeemed into another fake review for an app specified.
As a result, the amount of requirements-related information included in fake reviews could be influenced since some users of the portals might be app developers. 

% In-app purchases
For in-app purchases, we were unable to receive the offers programatically. Therefore, we could not compare the manually collected in-app purchases for apps affected by fake reviews against other app store in-app purchases. Also, we could not identify statistics which we could have used alternatively, or statistics on the monetization through ads.

% App Store
However, we decided to focus on the Apple App Store because of our prior experience with the technology and because this app store does not impose major API limitations to retrieve its data, e.g., compared to Google Play with limits the number of accessible reviews to 2,000 per app  and uses captchas.
% Dark patterns not analyzed
To have reliable results for the Apple App Store itself we crawled the largest dataset of about 62 million app reviews which has been analyzed so far to our knowledge. Thereby, we also avoid the App Sampling Problem for app store mining, described by Martin et al. \citep{Martin:2015:ASP:2820518.2820535}.

% Survey
The questionnaire we conducted with the paid review providers was hidden as a request for buying app reviews. We cannot assure that the responses only contain true statements. Therefore, we contacted several providers again after a few weeks using a different identity and communication channel, such as Skype. For providers we contacted again, their responses did not change.

% Human annotators
When manually labelling data, such as when finding agreements for potential matches between reviews within the fake reviews dataset and official reviews dataset, we used two human annotators which independently solved the task. In case of mismatches (3\%), we resolved the conflicts using a third human annotator. However, single reviews could have been mismatched.

% Effect Size
For statistical tests, in addition to reporting the p value we also calculated the effect size.
For t-tests we calculated the effect size using Cohen's d, that is the difference between means divided by the pooled standard deviation \citep{cohen1988spa}.
For Wilcoxon tests we calculated the r value, dividing the z distribution by the square root of the number of samples \citep{fritz2012effect}.
We report the effect size considering the following values, for Cohen's d (0.2 = small, 0.5 = medium, 0.8 = large) and for the correlation coefficient r (0.10 = small, 0.30 = medium, 0.50 = large).
In two cases, although a statistical difference was observed the effect size revealed that the magnitude between differences is near zero.
For these, the tests need to be repeated with additional samples (i.e., fake reviews) to show a statistical difference.

% Classifier
We want to stress that our classifier is only a first attempt to automatically identify fake reviews and not the main contribution of our paper. We wanted to verify if the features identified in our study are relevant for identifying fake reviews. 

The machine learning model could be overfitted. This may be due to the small amount of fake reviews. More fake reviews need to be collected to improve the results. We tried to minimize overfitting by using k-fold cross validation. To minimize the impact of randomly chosen data, we used another 8,000 randomly selected official reviews and were able to reproduce our results reported in the paper.

Moreover, we cannot ensure that all official reviews are non-fake reviews. As stated before, to provide a gold-standard dataset for regular reviews as well, all fake reviews must be known and removed which is the problem we are trying to solve in this paper. For this reason, we optimized the classifier for precision only. 

We leave the development of an advanced classifier for future research. Additional features, e.g., the emotion of users \citep{calefato2017emotxt, Martens:2017:EUA:3105556.3105559}, have to be analyzed to strengthen the results. For that we publicly share our gold-standard fake reviews dataset within our replication package.

% ------------------------------------------------------------------------------
% ------------------------------------------------------------------------------

\section{Related Work}
\label{sec:relwork}

User reviews are a valuable resource for decision making -- both to other users and developers. 
To our best knowledge no published studies on fake reviews for software products exist. 

A similar phenomena to fake reviews for software products, called web spamming, has been studied earlier when web pages began to compete for the rank within search engines' results.
This is comparable to apps competing within app stores today.
Web spamming is defined as the act of misleading search engines to rank pages higher than they deserve. 
Website operators edit their pages, e.g., by repeatedly adding specific terms that improve their ranking in search results \citep{gyongyi2005web}.
Based on this a definition for user-generated content emerged, called opinion spam.
The authors divide opinion spam into three categories, of which the first category are untruthful opinions.
These mislead readers and opinion mining systems by giving undeserved/unjust either positive reviews to promote or negative reviews to damage the reputation of a target object.
Untruthful opinions are also commonly known as fake reviews \citep{Jindal:2008:OSA:1341531.1341560}.
This definition has been refined several times, e.g., by adding that fake reviews are written by persons as if they were real customers \citep{Ott:2011:FDO:2002472.2002512}.

Recent research in other areas studied fake reviews, e.g., for products sold on Amazon, hotels rated on TripAdvisor, and businesses rated on Yelp.
Jindal and Liu \citep{Jindal:2008:OSA:1341531.1341560} first analyzed opinion spam. The authors analyze 5.8 million reviews and 2.14 million reviewers from Amazon to detect spam activities and present techniques to detect those. Due to the difficulty to create a fake reviews dataset, the authors used duplicate and near-duplicate reviews written by the same reviewers on different products. 
In our work we were able to extract data from fake review providers to achieve more reliable results. 

Ott et al.~\citep{Ott:2011:FDO:2002472.2002512} state that an increasing amount of user reviews is provided. Due to their value, platforms containing user reviews are becoming targets of opinion spam for potential monetary gain -- our work confirms this for app stores and provides further insight on the fake review offers and policies.
The authors focus on analyzing deceptive fake reviews, which are reviews that have been written to sound authentic, instead of disruptive fake reviews.
The authors highlighted that there are few sources for deceptive fake reviews. To overcome the issue they hired 400 humans using a crowd-sourcing platform to write fake hotel reviews. Their classifier integrates work from psychology and computational linguistics. It has an accuracy of nearly 90\% on the crowdsourced dataset. The authors showed that classifiers are better in recognizing deceptive fake reviews compared to humans which scored an accuracy of 61\% at most.

Feng et al. \citep{feng2012distributional} analyzed fake reviews on TripAdvisor and Amazon. They identified fake reviews based on the hypothesis that for a given domain a representative distribution of review rating scores exist which is distorted by fake reviews. The authors used an unsupervised learning approach to create a review dataset that is labeled automatically based on rating distributions. Using a statistical classifier trained on that dataset the authors were able to detect fake reviews with an accuracy of 72\%.

Mukherjee et al. \citep{Mukherjee2013WhatYF}, compared to existing studies, used real fake reviews instead of pseudo-fake reviews, e.g., generated using crowdsourcing platforms. Their dataset consists of fake reviews published on Yelp, filtered and marked by the platform itself. The authors used the supervised approach of Ott et al. \citep{Ott:2011:FDO:2002472.2002512} on their dataset an achieved a significantly lower accuracy of 67.8\%, compared to 89.6\%. The authors found that the word distribution of pseudo-fake reviews is different to the word distribution of real reviews. However, this does not apply for fake reviews within their dataset. Instead of using linguistic features, the authors suggest to use behavioral features. These include numeric values, such as the maximum number of reviews of a reviewer per day, or the review length. 
We followed this approach when designing our classifier and achieved encouraging results of up to 97\% precision.

Fake reviews have also been frequently discussed within the media.
Streitfeld \citep{Weblink:10015} reported that every fifth review submitted to Yelp is detected as dubious by its internal filters. Instead of removing dubious reviews, these are moved to the second page where they are read by less users. As fake reviews further increase, Yelp began a 'sting' campaign to publicly expose businesses buying fake reviews.

% ------------------------------------------------------------------------------
% ------------------------------------------------------------------------------

\section{Conclusion}
\label{sec:conclusion}

App reviews can be a valuable, unique source of information for software engineering teams reflecting the opinions and needs of actual users.  
Also potential users read through the reviews before deciding to download an app, similar to buying other products on the Internet.
Our work shows that part of app reviews in app stores are fake -- that is, they are incentivized and might not reflect spontaneous, unbiased opinions.

We analyzed the market of fake review providers and their fake reviewing strategies and found that developers buy reviews to relatively expensive prices of a few dollars or deal with reviews in exchange portals. 
Fake reviews are written to look authentic and are hard to recognize by humans. 
We identified differences between fake and official reviews.
We found that properties of the corresponding app and reviewer are most useful to determine if a review is fake. 
Based on the identified differences, we developed, trained, fine-tuned, and compared multiple supervised machine learning approaches. 
We found that the Random Forest classifier identifies fake reviews, given a proportional distribution of fake and regular reviews as reported in other domains, with a recall of 91\% and AUC/ROC value of 98\%. 
We publicly share our gold-standard fake reviews dataset to enable the development of more accurate classifiers to identify fake reviews.
Our work helps app store mining  researchers to sample apps and perform data cleaning to achieve more reliable results. 
Further, tools for app users and store operators can be built based on our findings to detect if app reviews are trustworthy and to take further actions against fake reviewers.

% ------------------------------------------------------------------------------

\section*{Acknowledgment}

This research was partially supported by the European Union Horizon 2020 project OpenReq under grant agreement no. 732463.

% ------------------------------------------------------------------------------
% ------------------------------------------------------------------------------

%%%%%%%%%%%%%%%%%%%%%%%%%%%%%%%%%%%%%%%%%%%%%%%%%%%%%%%%%%%%%%%%%%%

\bibliographystyle{spbasic}      % basic style, author-year citations
\bibliography{refs}   % name your BibTeX data base

\end{document}